\documentclass[aps]{revtex4}
\usepackage{amsmath,amssymb}
\usepackage{color,graphicx}
\newcommand{\be}{\begin{equation}}
\newcommand{\ee}{\end{equation}}
\newcommand{\calka}{\int\limits_a^b\int\limits_c^d}
\begin{document}
\title{Thermalization of  L\'{e}vy flights:  Path-wise picture in  2D}
\author{Mariusz  \.{Z}aba  and   Piotr Garbaczewski}
\affiliation{Institute of Physics, University of Opole, 45-052 Opole, Poland}
\begin{abstract}
We analyze two-dimensional (2D)  random  systems   driven by  a symmetric  L\'{e}vy stable noise  which, under
the sole influence of external (force) potentials   $\Phi (x) $,  asymptotically set down at   Boltzmann-type thermal  equilibria.
Such  behavior   is  excluded  within  standard  ramifications of  the Langevin approach to   L\'{e}vy flights. There,
the action of a conservative  force field  $\sim - \nabla \Phi (x)$    stands    for  an explicit reason
for   the  emergence of  an  asymptotic  invariant probability density function (pdf).  However the  latter cannot be represented  in the Boltzmann form
  $\rho _* (x) \sim \exp[- \Phi (x)]$ and the   thermal equilibrium concept appears to
 be alien to Langevin-modeled   L\'{e}vy flights.
In the present paper   we  address the   response of  L\'{e}vy noise  not to an external conservative  force field,  but directly
 to its   potential   $\Phi (x)$. That   is   explicitly  encoded    in  non-symmetric  jump transition rates  of the master equation
 for the  pdf    $\rho (x,t)$.
We  prescribe  a priori the  target  pdf   $\rho _*$    in the Boltzmann form $\sim \exp[- \Phi (x)]$ and next select the   L\'{e}vy  noise of interest.
Given suitable  initial data, this    allows  to infer a   reliable   path-wise  approximation to    a  true (albeit analytically beyond the reach)  solution of the  pertinent master equation,   with the property  $\rho (x,t)\rightarrow \rho _*(x)$  as time $t$  goes to infinity.
 No  explicit path-wise description   has  been  so far  devised for such  thermally  equilibrating   random motion.
 To reconstruct   random paths  of the   underlying  stochastic process  we   resort to numerical methods, where long jumps of the L\'{e}vy stable processes are statistically significant, but   are  truncated to become amenable to  simulation procedures.
 We  create  a  suitably modified version  of the time honored  Gillespie's algorithm,  originally  invented  in  the
  chemical kinetics context. A  statistical  analysis  of    generated    sample trajectories    allows   us to
 infer   a surrogate   pdf  dynamics  which   consistently  sets down at a pre-defined target pdf.
   We pay special  attention to the response of the  2D  Cauchy noise    to an exemplary   locally  periodic  "potential landscape"   $\Phi (x), x\in R^2$.
\end{abstract}
  \maketitle

\section{Introduction}

Various   random processes  in real physical systems  admit  a  simplified   description based on stochastic differential equations.
Then,  there is a routine  passage  procedure  from microscopic   random variables to macroscopic (statistical  ensemble, mean field) data,    like e.g.   the   time evolution of an associated probability density function (pdf)  which  is  a solution of
  a deterministic  transport equation. A paradigm    example  is    so-called Langevin modeling of diffusion-type and jump-type processes.
The presumed microscopic model of  random dynamics   is provided by the Langevin (stochastic)  equation,   which   additively  decomposes into
a    (Newtonian by  origin) drift   and   purely random (perturbing noise)   term.    Its  direct  consequence is   the  Fokker-Planck   equation  for
  an associated probability density function (pdf), \cite{risken} and \cite{fogedby}.

As a necessary pre-requisite  for    our  further discussion, let us discuss  a   transformation of  the  Fokker-Planck equation
into   the  Schr\"{o}dinger-type (generalized diffusion) equation, often employed   in the theoretical  framework of the Brownian motion.
 Here,  the   Langevin  equation,  the  induced  Fokker-Planck equation  and its  Schr\"{o}dinger -type    image     are dynamically  equivalent and
 describe    the   same  diffusion-type process.   This is not the case  if one   turns over to jump-type processes.

 For clarity of arguments, let us consider the  Langevin equation for a one-dimensional   diffusion
process in an  external conservative  force field $F(x) = - dV(x)/dx$  in the   form
${\frac{dx}{dt}} = F(x) + \sqrt{2\nu } b(t)$,    where $b(t)$ stands for the normalized white noise: $\langle b(t)\rangle =0$,
 $\langle b(t')b(t)\rangle= \delta (t-t')$ and the  mass parameter is scaled away.
 The corresponding Fokker-Planck equation   for the probability density  function  $\rho (x,t)$   reads
\be
 \partial _t\rho = \nu \Delta \rho  - \nabla ( F\, \rho)
\ee
  and, in the confining regime,  is known to enforce  the existence of an asymptotic invariant pdf,   $\rho (x,t) \rightarrow \rho _*(x)$  as  $t\rightarrow \infty $,     in the explicit  Boltzmann form  $\exp (- \Phi  /2)$, where  $\Phi (x) = V(x)/\nu $.

 By  means of a standard substitution  $\rho (x,t) = \psi (x,t) \exp[- V(x)/2\nu]$, \cite{risken},  the Fokker-Planck equation   can be
 transformed into  a  generalized diffusion equation for an auxiliary function $\psi (x,t)$. This   Schr\"{o}dinger-type
  equation (no imaginary unit $i$) reads
\be
    \partial _t \psi = \nu \Delta \psi - {\cal{V}} (x) \psi
 \ee
 where $ {\cal{V}} (x) = {\frac{1}2}\left( {\frac{F^2}{2\nu }} + \nabla  F\right)$  and $F= F(x)$.

By   reintroducing  a  normalization constant     (divide   and multiply by a suitable  number $Z^{1/2}$
   in the factorization formula for $\rho (x,t)$), we
  can   rewrite $\rho (x,t)$ in the   form   $\rho (x,t) = \Psi (x,t) \rho _*^{1/2}(x)$, where
   $\rho _*^{1/2} =Z^{-1/2}\,  \exp (- \Phi  /2) $   while  $\Psi = Z^{1/2} \psi $.  Clearly, $\Psi (x,t) \rightarrow \rho _*^{1/2}(x)$
    as  $t$ goes to infinity.  Moreover, we can rewrite  the semigroup potential    as  follows:
      $ {\cal{V}}    (\nu \, \Delta \rho _*^{1/2})/\rho ^{1/2}_*$.

The   transformation of (1) into (2)     cannot be    adopted  to   L\'{e}vy jump-type processes,   where  the Langevin
and  Schr\"{o}dinger-type  (semigroup)  modeling   are   known to be incompatible.   Moreover,   the Eliazar-Klafter  no go  statement, \cite{klafter},
 disconnects the Langevin-modeled    Fokker-Planck equation  for any L\'{e}vy-stable noise
\begin{equation}
\dot{x}= b(x)  + A^{\mu }(t)  \Rightarrow
 \partial _t\rho = -\nabla (b\cdot \rho ) - \lambda |\Delta |^{\mu /2}\rho \, .  \label{fraceq}
 \end{equation}
from the very notion of the Boltzmann   thermal equilibrium.

However,   the  thermal equilibrium  notion  remains a valid concept  within an  immediate  L\'{e}vy  transcript   of the
 semigroup dynamics (2) (e.g.  replace $\nu \Delta $ by $-\lambda |\Delta |^{\mu/2}$):
\begin{equation}
  \partial _t\Psi = -  \lambda |\Delta |^{\mu /2} \Psi
- {\cal{V}} \Psi  \, ,  \label{semig}
\end{equation}
 see e.g.  \cite{olk,gar1}, where we assume that  $\Psi (x,t)$  asymptotically  sets down at a square root $\rho _*^{1/2} (x)$
  of  a well defined pdf $\rho _*$.  The   semigroup  potential  ${\cal{V}}(x)$
 follows from the  compatibility condition:
  \begin{equation}
 {\cal{V}}  =   -\lambda\,  {\frac{|\Delta |^{\mu /2}\,  \rho ^{1/2}_*}{\rho ^{1/2}_*}} \, .  \label{pot}
  \end{equation}

In  this   particular  context, while adopting a multiplicative  decomposition of the time-dependent pdf  $\rho (x,t) \rightarrow \rho _*(x)$:
  \begin{equation}
    \rho (x,t) = \Psi (x,t) \rho _*^{1/2}(x)\, ,  \label{def}
    \end{equation}
      a  novel   fractional  generalization  of the   Fokker-Planck equation    governing the time evolution of
  $\rho (x,t) $    has   been introduced in Refs. \cite{brockmann}-\cite{belik}, see also
  \cite{gar1,gar2,gar},   to handle systems that are randomized  by symmetric  L\'{e}vy-stable  drivers and may asymptotically set
   down at Boltzmann-type equilibria   under the influence of external potentials (thus not Newtonian forces anymore).

The pertinent Fokker-Planck type equation, whose  origin has been discussed  before   in a number of papers, \cite{brockmann}-\cite{gar},
 has the  familiar  master equation form, presently  reproduced   in the explicit 2D form:
\be
\partial_t\rho(x,y)=\iint\limits_A [w_\phi(x,y|u,v)\rho(u,v)-w_\phi(u,v|x,y)\rho(x,y)]\nu_\mu(x,y,dx,dy),\label{l1}
\ee
Here,   anticipating    the  effectiveness of     numerical routines  to be described in below,  from the start   we   impose    cut-offs upon the     size of jumps
to be  accounted for   during the simulations (large     jump sizes  remain  statistically significant if view of the involved    L\'{e}vy distribution)
\be
\begin{split}
&A=\{(x,y)\in\mathbb{R}^2;\quad {\varepsilon_1^x\leqslant|x-u|\leqslant \varepsilon_2^x}\,\wedge\,
{\varepsilon_1^y\leqslant|y-v|\leqslant \varepsilon_2^y}\},\\
&w_\phi(x,y|u,v)=\exp[(\Phi(u,v)-\Phi(x,y))/2],\label{l2}
\end{split}
\ee
and   the  L\'{e}vy  measure  $\nu_\mu(x,y,dx,dy)$    in $\mathbb{R}^2$  is given by
\be
\nu_\mu(x,y,dx,dy)=\frac{2^\mu\Gamma((2+\mu)/2)}{\pi|\Gamma(-\mu/2)|}\frac{1}{(x^2+y^2)^{(2+\mu)/2}}dx\,dy=\widetilde{\nu}_\mu(x,y)dx\,dy,\qquad \mu\in(0,2).\label{l3}
\ee
It is the quantity    $w_\phi(x,y|u,v)\widetilde{\nu}_\mu(x,y)$   which has an interpretation of  the jump transition rate    from the point  $(u,v)  \in R^2$   to another point   $(x,y) \in R^2$.
 The  potential   function   $\Phi(x,y)$  can be chose quite arbitrarily. However, we  need to secure a  $L^1(R^2)$   normalization   of the target pdf  $\rho_*(x,y)  \sim  \exp(-\Phi(x,y))$.
 We note that  $\exp(-\Phi(x,y))$  becomes  a genuine   stationary solution of   Eq.~ (\ref{l1})  once we let  $\varepsilon_1^{x,y} \to 0$   and  $\varepsilon_2^{x,y} \to\infty$.

\begin{figure}[h]
\begin{center}
%\centering
\includegraphics[width=45mm,height=45mm]{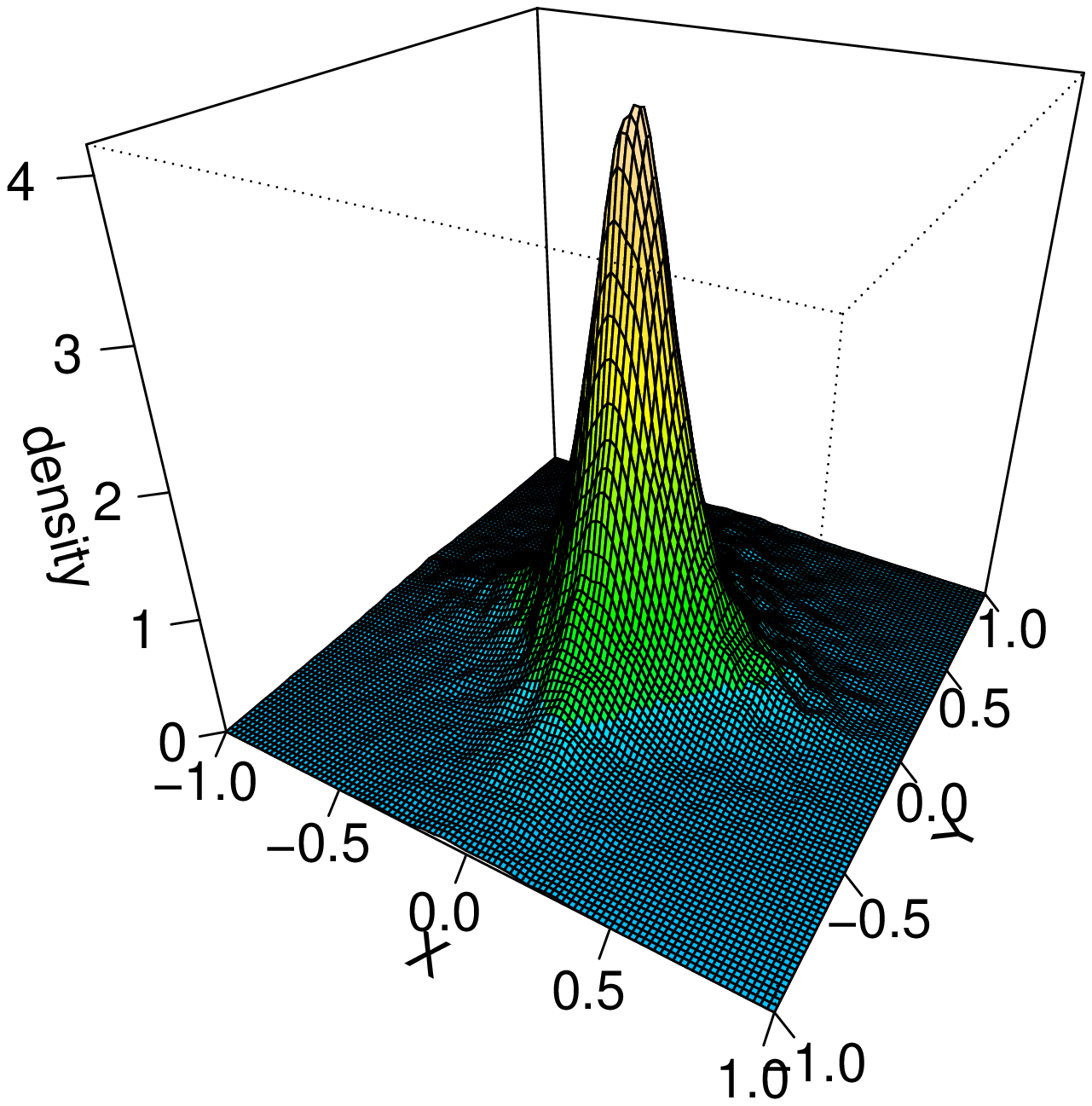}
\includegraphics[width=45mm,height=45mm]{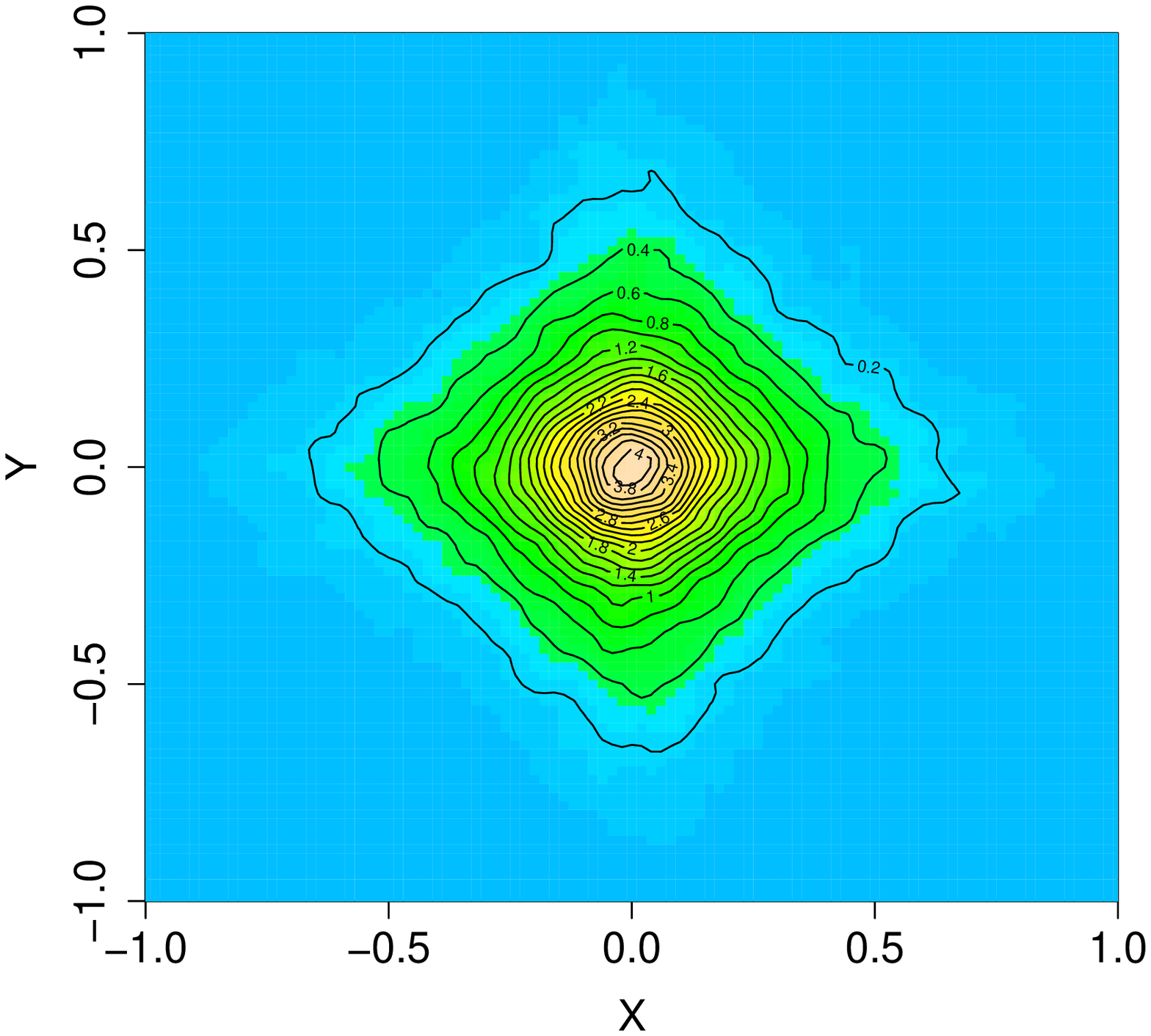}\\
\includegraphics[width=45mm,height=45mm]{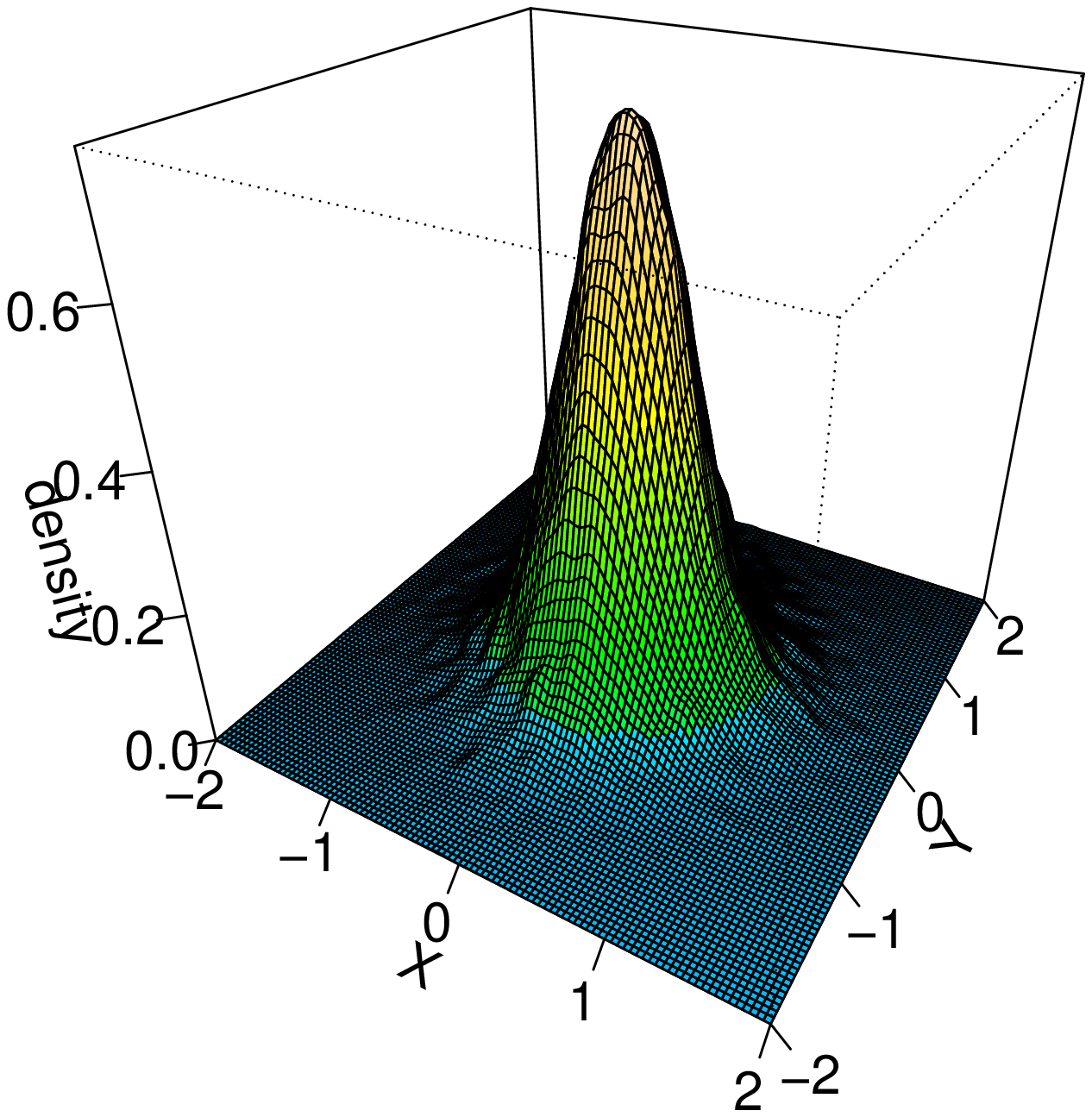}
\includegraphics[width=45mm,height=45mm]{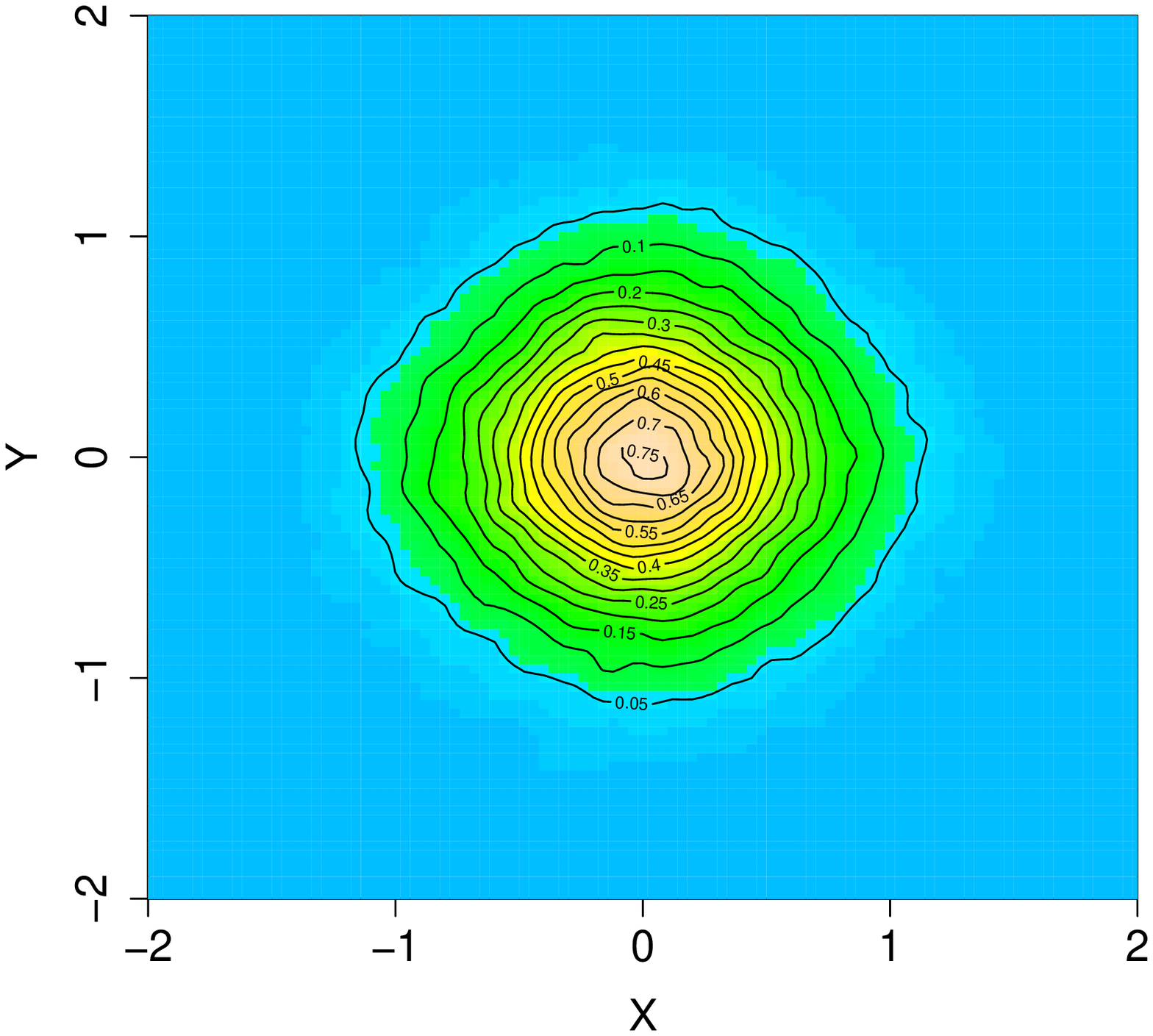}\\
\includegraphics[width=45mm,height=45mm]{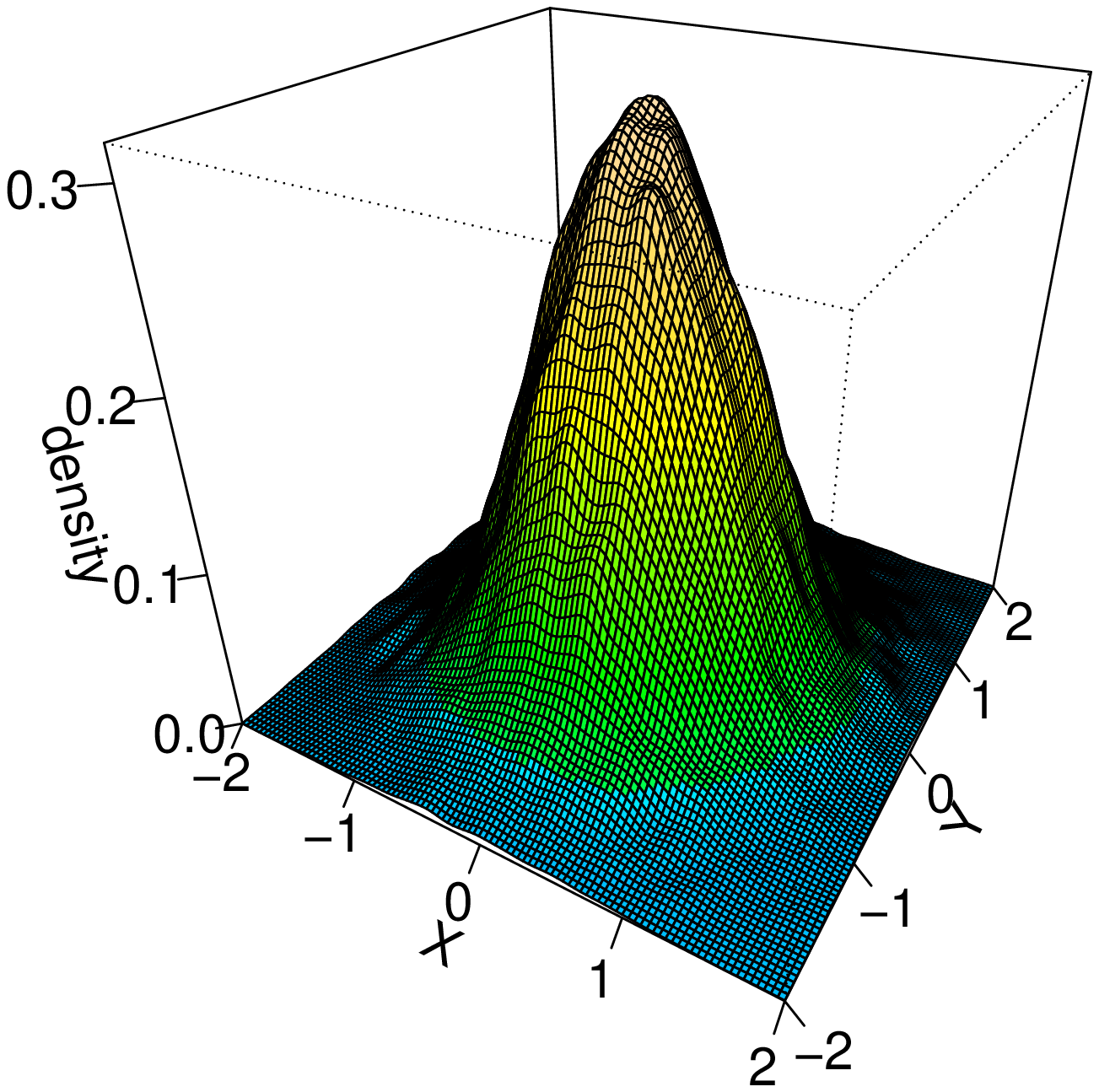}
\includegraphics[width=45mm,height=45mm]{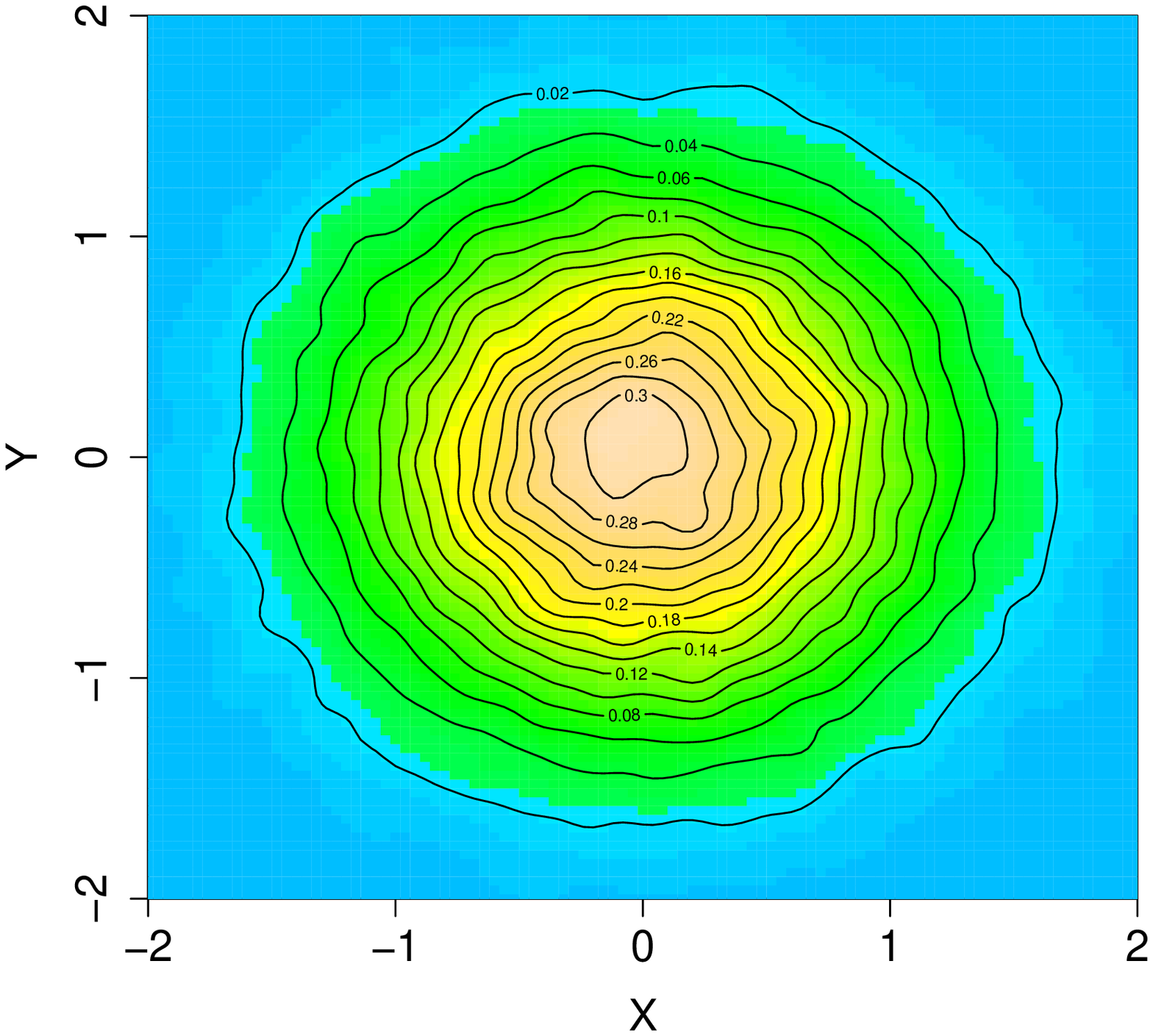}\\
\includegraphics[width=45mm,height=45mm]{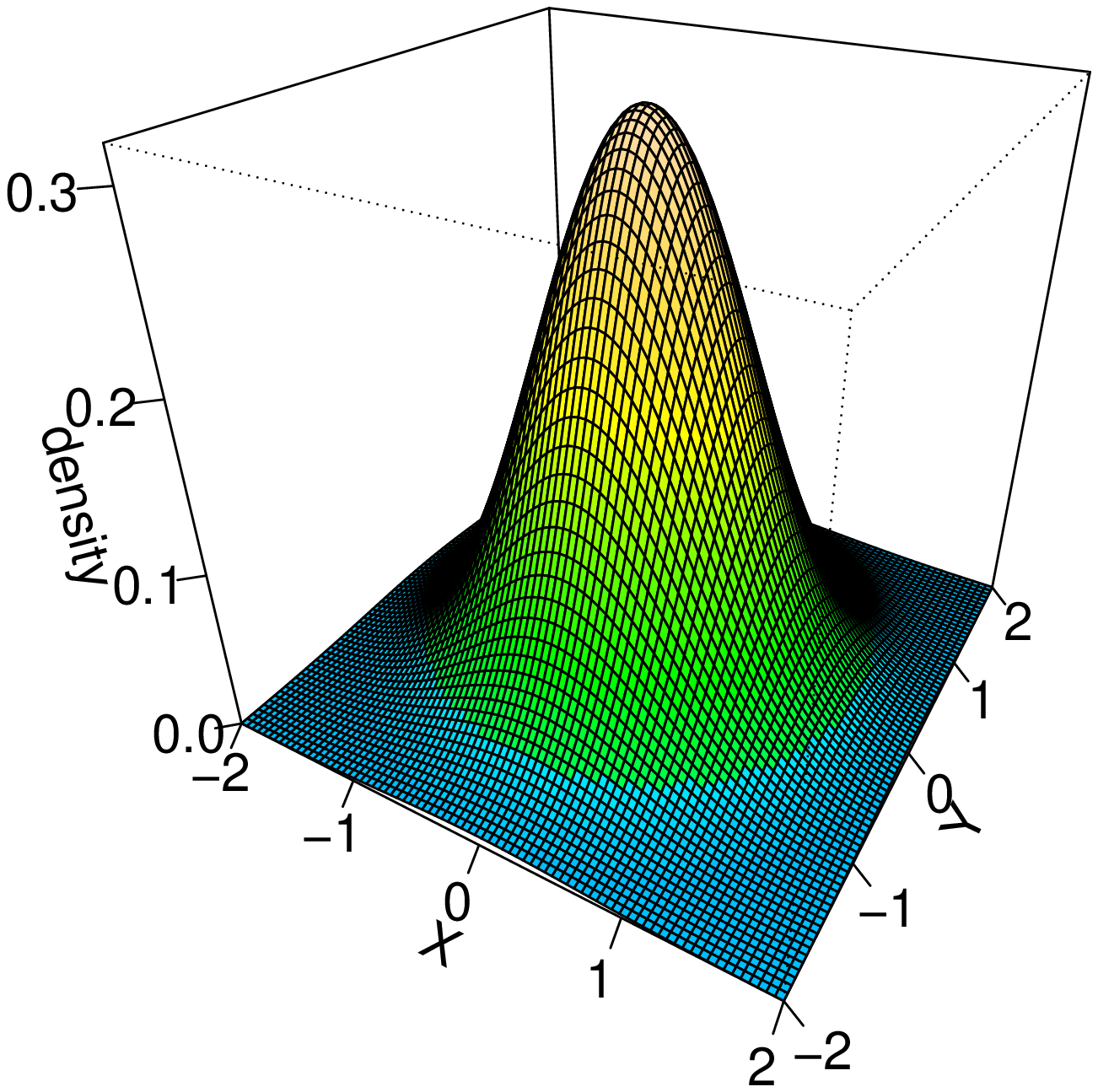}
\includegraphics[width=45mm,height=45mm]{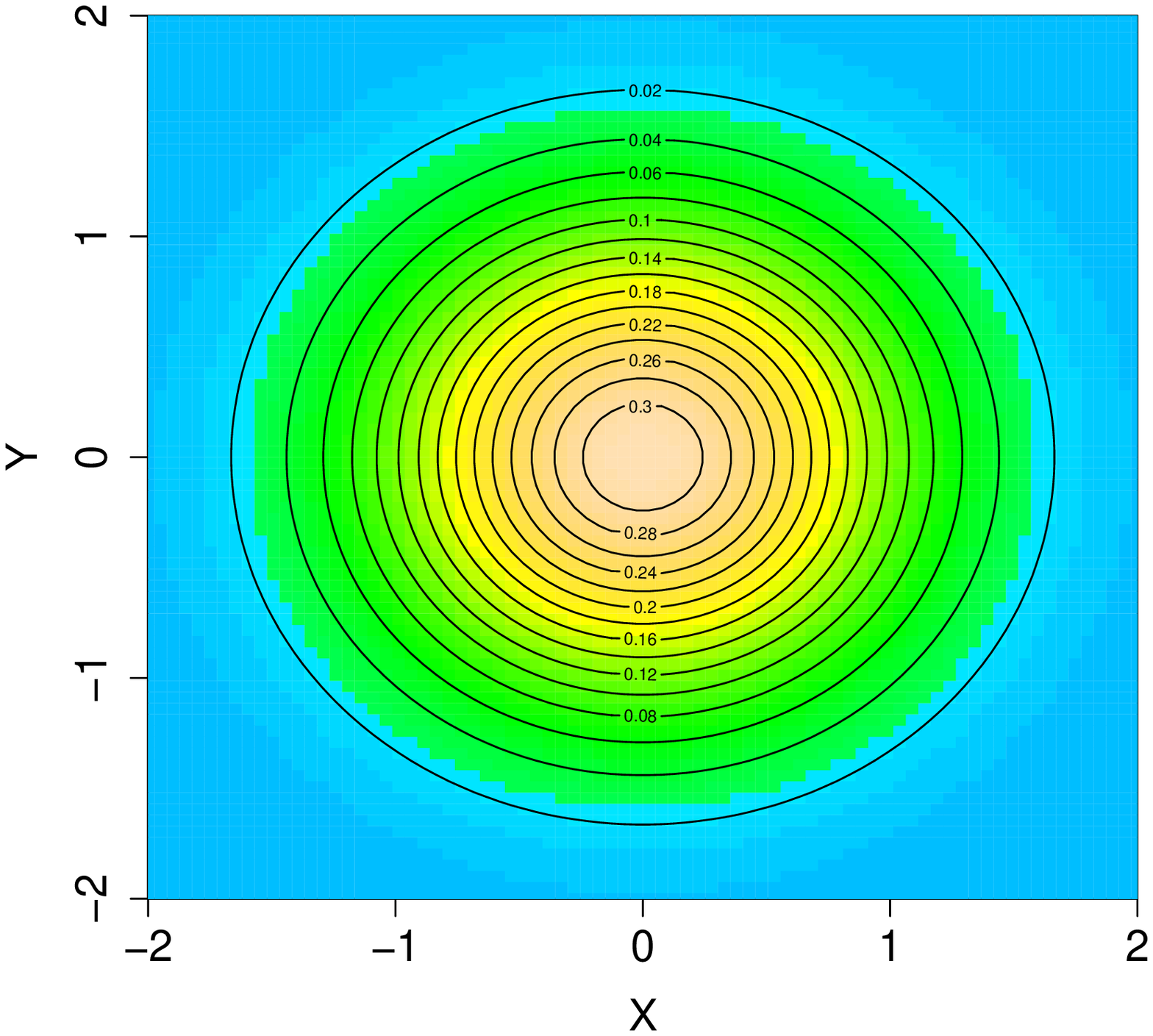}
\end{center}
\caption{ Gaussian  target;  statistics of    $100 000$    random paths: surrogate pdf evolution   and its OXY projection  for  a) $t=0.2$, b) $t=0.7$, c) $t=15$. The subfigure   d)   refers to the  asymptotic    pdf  (\ref{l8}).
All trajectories were
started form  the origin  $(0,0)$.}
%\end{center}
\end{figure}

\section{Gillespie's  algorithm: Fine tuning   in 2D.}

The Gillespie's algorithm has been originally constructed to give a dynamical picture of  a finite chain of  chemical reactions, \cite{gillespie,gillespie1}.  There,  switches between different  chemical reaction channels can be
 re-interpreted as   jumps between points in a finite state space.  Since the number of allowed chemical channels is finite,   a serious   modification of an original algorithm must be created to account
 for    very large   (virtually infinite) state space  we need to consider in connection with L\'{e}vy flights.  As an  example, a  jump process analog of chemical reaction channels could  comprise
 (take $R^1$ for a while)    all jumps   form a fixed point $x_0$ to any point in  the set $[x_0-\epsilon _2,x_0 - \epsilon _1]\cup [x_0 + \epsilon _1, x_0 +\epsilon _2]$.
 Since we aim at numerical simulations, it is obvious that all admissible jumps cannot  be computer-generated. Therefore, we  must restrict considerations to a large number of
 suitably selected  "representative" jumps of a truncated jump process   ruled by the truncated L\'{e}vy distribution of jumps.

Basic tenets of the modified  Gillespie's  algorithm,   fine tuned to account for L\'{e}vy jumps in $R^1$ have been described in detail elsewhere, \cite{zaba}.
Presently, we  shall give a brief outline of the algorithm  that  is capable to  account   for   L\'{e}vy flights     in $R^2$.   We mimic  previously  devised 1D steps,  \cite{zaba},   while  adopting them to the
 2D situation.

 Eq.  (\ref{l1})  can be rewritten as follows
\be
\partial_t\rho(x,y)=\iint\limits_{A'} C_\mu[w_\phi(x,y|\xi+x,\eta+y)\rho(\xi+x,\eta+y)-w_\phi(\xi+x,\eta+y|x,y)\rho(x,y)]d\xi\,d\eta,\label{l4}
\ee
where
\be
\begin{split}
&A'=\{(\xi,\eta)\in\mathbb{R}^2;\quad {\varepsilon_1^x\leqslant|\xi|\leqslant \varepsilon_2^x}\,\wedge\,
{\varepsilon_1^y\leqslant|\eta|\leqslant \varepsilon_2^y}\},\\
&C_\mu=\frac{2^\mu\sin(\pi\mu/2)[\Gamma((\mu+2)/2)]^2}{\pi^2(\xi^2+\eta^2)^{(\mu+2)/2}},\label{l5}
\end{split}
\ee
The algorithm outline reads as follows:\\
(i) Set time  $t=0$  and the point of origin for jumps, $(x_0,y_0)\in R^2$.\\
(ii)  Create the set of all admissible jumps from $(x_0,y_0)$ to $(x_0 + \xi , y_0+ \eta )$, compatible with the transition  rate  $w_\phi(x,y|x+\xi,y+ \eta)$.\\
(iii)  Evaluate
\be
\begin{split}
&W_1(x_0,y_0)=\int\limits_{-\varepsilon_2^x}^{-\varepsilon_1^x}\int\limits_{-\varepsilon_2^y}^{-\varepsilon_1^y}C_\mu w_\phi(\xi+x_0,\eta+y_0|x_0,y_0)d\xi\,d\eta, \quad W_2(x_0,y_0)=\int\limits_{-\varepsilon_2^x}^{-\varepsilon_1^x}\int\limits_{\varepsilon_1^y}^{\varepsilon_2^y}C_\mu w_\phi(\xi+x_0,\eta+y_0|x_0,y_0)d\xi\,d\eta,\\
&W_3(x_0,y_0)=\int\limits_{\varepsilon_1^x}^{\varepsilon_2^x}\int\limits_{-\varepsilon_2^y}^{-\varepsilon_1^y}C_\mu w_\phi(\xi+x_0,\eta+y_0|x_0,y_0)d\xi\,d\eta,  \quad
W_4(x_0,y_0)=\int\limits_{\varepsilon_1^x}^{\varepsilon_2^x}\int\limits_{\varepsilon_1^y}^{\varepsilon_2^y}C_\mu w_\phi(\xi+x_0,\eta+y_0|x_0,y_0)d\xi\,d\eta, \label{l6}
\end{split}
\ee
and  $W(x_0,y_0)=W_1(x_0,y_0)+W_2(x_0,y_0)+W_3(x_0,y_0)+W_4(x_0,y_0)$.\\
(iv) Using a random number generator draw  $p_x,p_y\in[0,1]$  from a uniform distribution \\
(v) Employing   above $p_x$ and $p_y$ and the identities
\scriptsize
\be
\left\{
  \begin{array}{ll}
    \int\limits_{-\varepsilon_2^x}^{b_x}\left(\int\limits_{-\varepsilon_2^y}^{-\varepsilon_1^y}C_\mu w_\phi(\xi+x_0,\eta+y_0|x_0,y_0)d\eta+
    \int\limits_{\varepsilon_1^y}^{\varepsilon_2^y}C_\mu w_\phi(\xi+x_0,\eta+y_0|x_0,y_0)d\eta\right)d\xi =p_x W(x_0,y_0), & \hbox{$p_x<W_{12}(x_0,y_0)/W(x_0,y_0)$;} \\
    W_{12}(x_0,y_0)+\int\limits_{\varepsilon_1^x}^{b_x}\left(\int\limits_{-\varepsilon_2^y}^{-\varepsilon_1^y}C_\mu w_\phi(\xi+x_0,\eta+y_0|x_0,y_0)d\eta+
    \int\limits_{\varepsilon_1^y}^{\varepsilon_2^y}C_\mu w_\phi(\xi+x_0,\eta+y_0|x_0,y_0)d\eta\right)d\xi=p_x W(x_0,y_0), & \hbox{$p_x\geqslant W_{12}(x_0,y_0)/W(x_0,y_0)$;}\\
    \int\limits_{-\varepsilon_2^y}^{b_y}\left(\int\limits_{-\varepsilon_2^x}^{-\varepsilon_1^x}C_\mu w_\phi(\xi+x_0,\eta+y_0|x_0,y_0)d\xi+
    \int\limits_{\varepsilon_1^x}^{\varepsilon_2^x}C_\mu w_\phi(\xi+x_0,\eta+y_0|x_0,y_0)d\xi\right)d\eta =p_y W(x_0,y_0), & \hbox{$p_y<W_{13}(x_0,y_0)/W(x_0,y_0)$;} \\
    W_{13}(x_0,y_0)+\int\limits_{\varepsilon_1^y}^{b_y}\left(\int\limits_{-\varepsilon_2^x}^{-\varepsilon_1^x}C_\mu w_\phi(\xi+x_0,\eta+y_0|x_0,y_0)d\xi+
    \int\limits_{\varepsilon_1^x}^{\varepsilon_2^x}C_\mu w_\phi(\xi+x_0,\eta+y_0|x_0,y_0)d\xi\right)d\eta=p_y W(x_0,y_0), & \hbox{$p_y\geqslant W_{13}(x_0,y_0)/W(x_0,y_0)$,}\label{l7}
  \end{array}
\right.
\ee
\normalsize
where  $W_{12}(x_0,y_0)=W_1(x_0,y_0)+W_2(x_0,y_0)$ i $W_{13}(x_0,y_0)=W_1(x_0,y_0)+W_3(x_0,y_0)$, find   $b_x$ and $b_y$ corresponding to the "transition channel"  $(x_0,y_0)\rightarrow  (b_x,b_y)$ .\\
(vi) Draw a   new number  $q\in(0,1)$   from a uniform distribution.\\
(vii)  Reset time label s $t=t+\Delta t$  where : $\Delta t=-\ln q/W(x_0,y_0)$.\\
(viii)   Reset     $(x_0,y_0)$ to a new  location      $(x_0\doteq x_0+b_x,  y_0 \doteq  y_0+b_y)$, considered as a reference point for  the iterative procedure.\\
(ix) Return to step (ii) and repeat the procedure anew.\\\\
\textbf{Remark:}  All precautions respected in 1D  need to be respected in 2D as well  (c.f. Comment 1 in \cite{zaba}).  The following jump size bounds   (integration boundaries)   were adopted for
exemplary numerical procedures to be described in below: $\varepsilon_1^x=\varepsilon_1^y=\varepsilon_1=0.001$ i $\varepsilon_2^x=\varepsilon_2^y=\varepsilon_2=1$, provided one keeps in memory our convention not to
   reproduce the  $x,y$ index in  $\varepsilon_{1,2}^{x,y}$  anymore.
\begin{figure}[h]
\begin{center}
%\centering
\includegraphics[width=45mm,height=45mm]{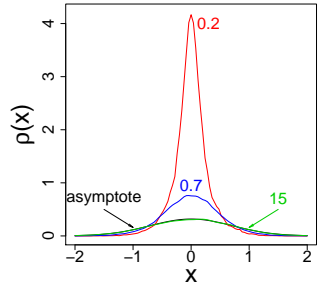}\\
\end{center}
\caption{ $y=0$ projection of the  previous  pdf data  ($100 000$ trajectories)  for running time instants  a) $t=0.2$, b) $t=0.7$, c) $t=15$, d)  a projection of the target pdf is depicted.}
%\end{center}
\end{figure}

\begin{figure}[h]
\begin{center}
\begin{minipage}[b]{45mm}
\centering
\includegraphics[width=45mm]{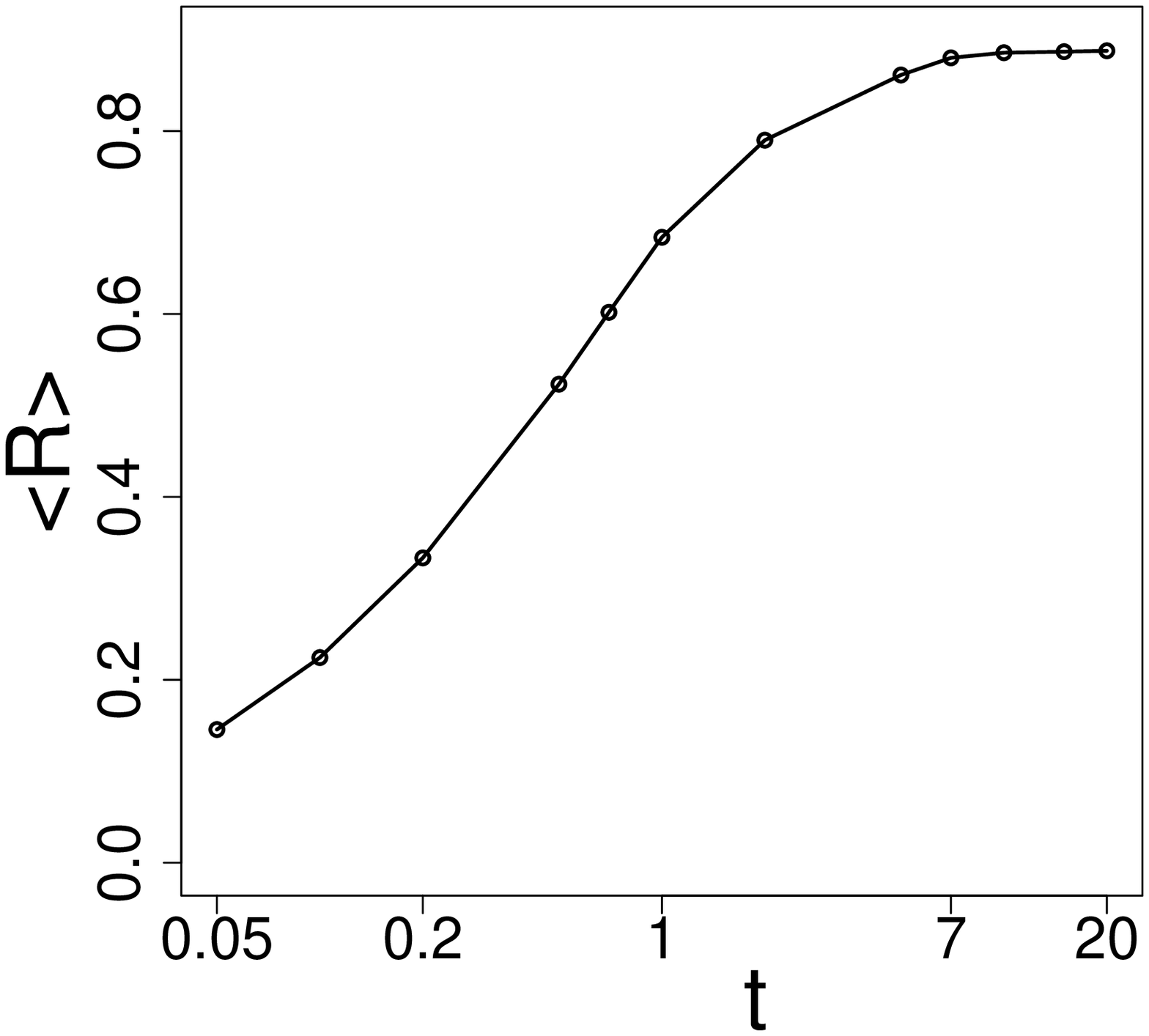}\\\textit{a)}
\end{minipage}
\begin{minipage}[b]{45mm}
\centering
\includegraphics[width=45mm]{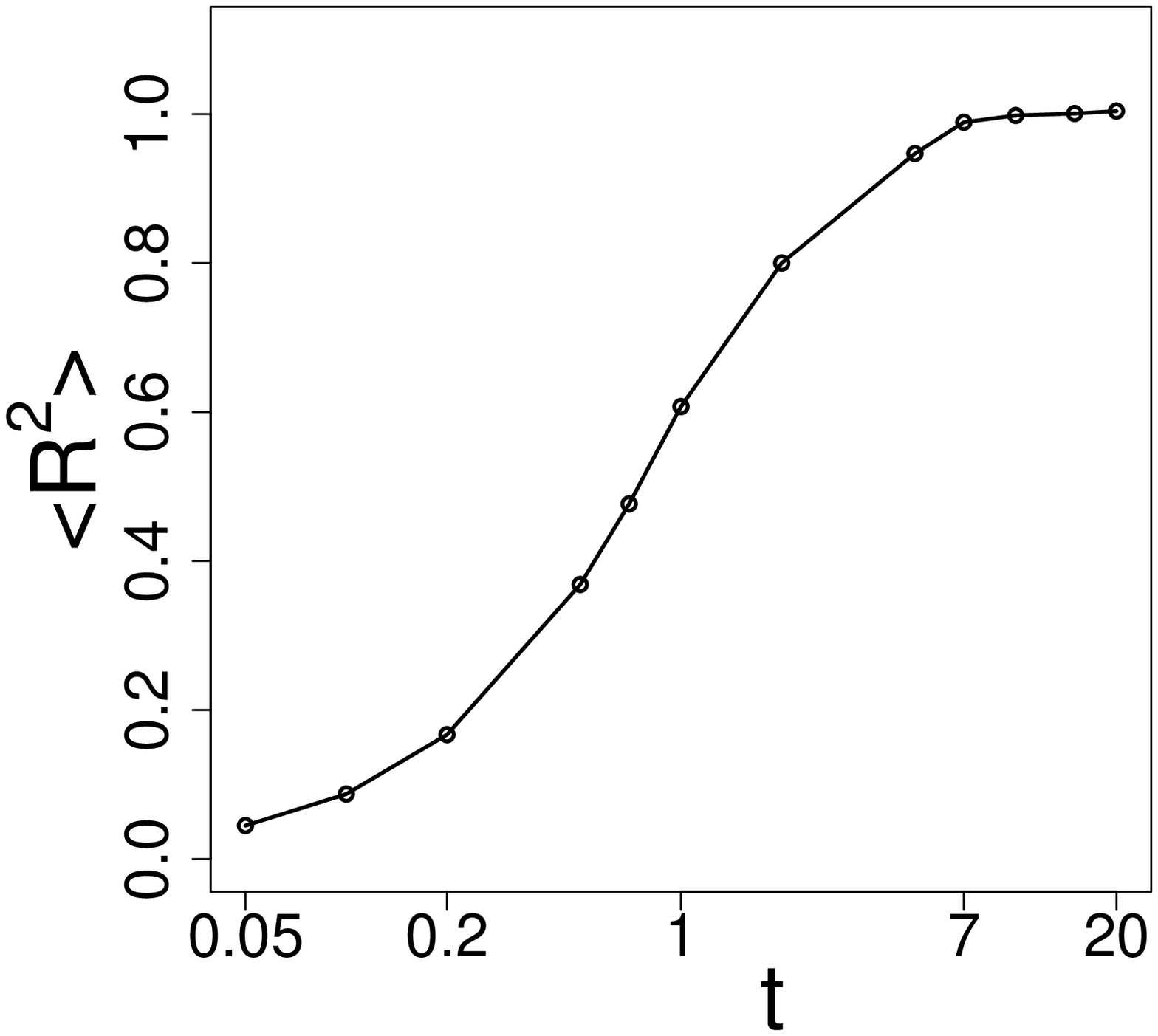}\\\textit{b)}
\end{minipage}
\caption{ Time evolution  of  a) the first moment  (average  $r=\sqrt{x^2 +y^2}$ value) , b)  second moment  (mean value of  $r^2$). All  $100 000$  trajectories originate from  $(0,0)$.}
\end{center}
\end{figure}

\section{Statistics of random paths   in 2D:   Case studies  of  the pdf evolution.}

Our  main  purpose   in the present section is to analyze  the response of     free  L\'{e}vy   noise   to   generic  external potentials. That has been
 previously found to be encoded in  jump  transition  rates. Now   we   address the same problem in  a path-wise fashion.  We shall  faithfully  follow the outlined  random path generation procedure.
Once suitable path ensemble data  are collected,    we shall  verify whether
statistical  (ensemble)   features of generated random trajectories  are  compatible with those  predicted by the    master equation  (\ref{l1}).
 This   includes a control  of  an asymptotic behavior   $\rho (x,t) \to \rho_*(x)$  with $x\in R^2$,  when  $t\to\infty$.

\subsection{Harmonic confinement: Gaussian target}

Let us prescribe  an asymptotic invariant pdf   $\rho _* $    to be in a   2D  Gaussian form:
\be
\rho_*(x,y)=\frac{1}{\pi}e^{-x^2-y^2}.\label{l8}
\ee
As an exemplary source of random noise we assume the  Cauchy  driver,  i.e.    2D   L\'{e}vy-stable   noise with the  stability index    $\mu=1$.
Accordingly, the jump transition rate reads:
\be
C_1\, w_\phi(\xi+x,\eta+y|x,y)=\frac{1}{2\pi}\frac{e^{-\xi^2/2-x\xi-\eta^2-y\eta}}{(\xi^2+\eta^2)^{3/2}}.\label{l9}
\ee

To generate sample paths of the process, we need first to evaluate   integrals   of  (\ref{l9})  over suitable integration  volumes.    If  the  integration volume
comprises   $x,y$   which are close to the  jump size  boundaries  $\pm\varepsilon_1$,  one develops a numerator of an expression into  Taylor series  and  keeps terms up to the quadratic one.
   Errors induced by such approximation procedure are marginal.   On the other hand, if  $x,y$  are far away from  $\pm\varepsilon_1$  integrals  (\ref{l9}) are amenable to
    standard evaluation methods (like e.g.   Simposon's one).    To be more explicit in this respect, let  $\varepsilon_{12}=0.05$. If
  $|a|,|b|,|c|,|d|\leqslant \varepsilon_{12}$, then
\be
\calka\frac{e^{-\xi^2/2-x\xi-\eta^2-y\eta}}{(\xi^2+\eta^2)^{3/2}}d\xi\,d\eta\thickapprox
\calka\frac{(1-x\xi+\frac{x^2-1}{2}\xi^2)(1-y\eta+\frac{y^2-1}{2}\eta^2)}{(\xi^2+\eta^2)^{3/2}}d\xi\,d\eta\thickapprox\sum\limits_{i=1}^6\mathbb{I}_i,\label{l10}
\ee
where
\be
\begin{split}
&\mathbb{I}_1=\calka\frac{d\xi\,d\eta}{(\xi^2+\eta^2)^{3/2}}=\int\limits_a^b\left(\frac{d}{\xi^2\sqrt{\xi^2+d^2}}-\frac{c}{\xi^2\sqrt{\xi^2+c^2}}\right)d\xi=
-\frac{\sqrt{b^2+d^2}}{bd}+\frac{\sqrt{a^2+d^2}}{ad}+\frac{\sqrt{b^2+c^2}}{bc}-\frac{\sqrt{a^2+c^2}}{ac},\\
&\mathbb{I}_2=-\calka\frac{x\xi d\xi\,d\eta}{(\xi^2+\eta^2)^{3/2}}=x\ln\left(\frac{(d+\sqrt{b^2+d^2})(c+\sqrt{a^2+c^2})}{(c+\sqrt{b^2+c^2})(d+\sqrt{a^2+d^2})}\right),\\
&\mathbb{I}_3=-\calka\frac{y\eta d\xi\,d\eta}{(\xi^2+\eta^2)^{3/2}}=y\ln\left(\frac{(b+\sqrt{b^2+d^2})(a+\sqrt{a^2+c^2})}{(a+\sqrt{a^2+d^2})(b+\sqrt{b^2+c^2})}\right),\\
&\mathbb{I}_4=\calka\frac{x y\xi\eta d\xi\,d\eta}{(\xi^2+\eta^2)^{3/2}}=x y(\sqrt{a^2+d^2}+\sqrt{b^2+c^2}-\sqrt{b^2+d^2}-\sqrt{a^2+c^2}),\\
&\mathbb{I}_5=\calka\frac{x^2-1}{2}\frac{\xi^2d\xi\,d\eta}{(\xi^2+\eta^2)^{3/2}}=
\frac{x^2-1}{2}\left(d\ln\left(\frac{b+\sqrt{b^2+d^2}}{a+\sqrt{a^2+d^2}}\right)-c\ln\left(\frac{b+\sqrt{b^2+c^2}}{a+\sqrt{a^2+c^2}}\right)\right),\\
&\mathbb{I}_6=\calka\frac{y^2-1}{2}\frac{\eta^2d\xi\,d\eta}{(\xi^2+\eta^2)^{3/2}}=
\frac{y^2-1}{2}\left(b\ln\left(\frac{d+\sqrt{b^2+d^2}}{c+\sqrt{b^2+c^2}}\right)-a\ln\left(\frac{d+\sqrt{a^2+d^2}}{c+\sqrt{a^2+c^2}}\right)\right).\label{l11}
\end{split}
\ee
Numerical routines were  written  in terms of C-codes, \cite{available}.

In Fig. 1 we have depicted  the statistical data inferred from  100 000  trajectories, for three running time instants $t=0.2$, $t=0.7$, $t=15$, together with the asymptotic expression  (\ref{l8}).
The right-hand-side column depicts  projections of those data upon the  the  $OXY$  plane. A substanitial   increase of the  analyzed  enesemble data (like e.g. $300 000$, $500  000$  or $10^6$) is merely a matter of
the simulation time span and adds nothing inspiring to the obtained behavior.
\begin{figure}[h]
\begin{center}
\centering
\includegraphics[width=45mm,height=45mm]{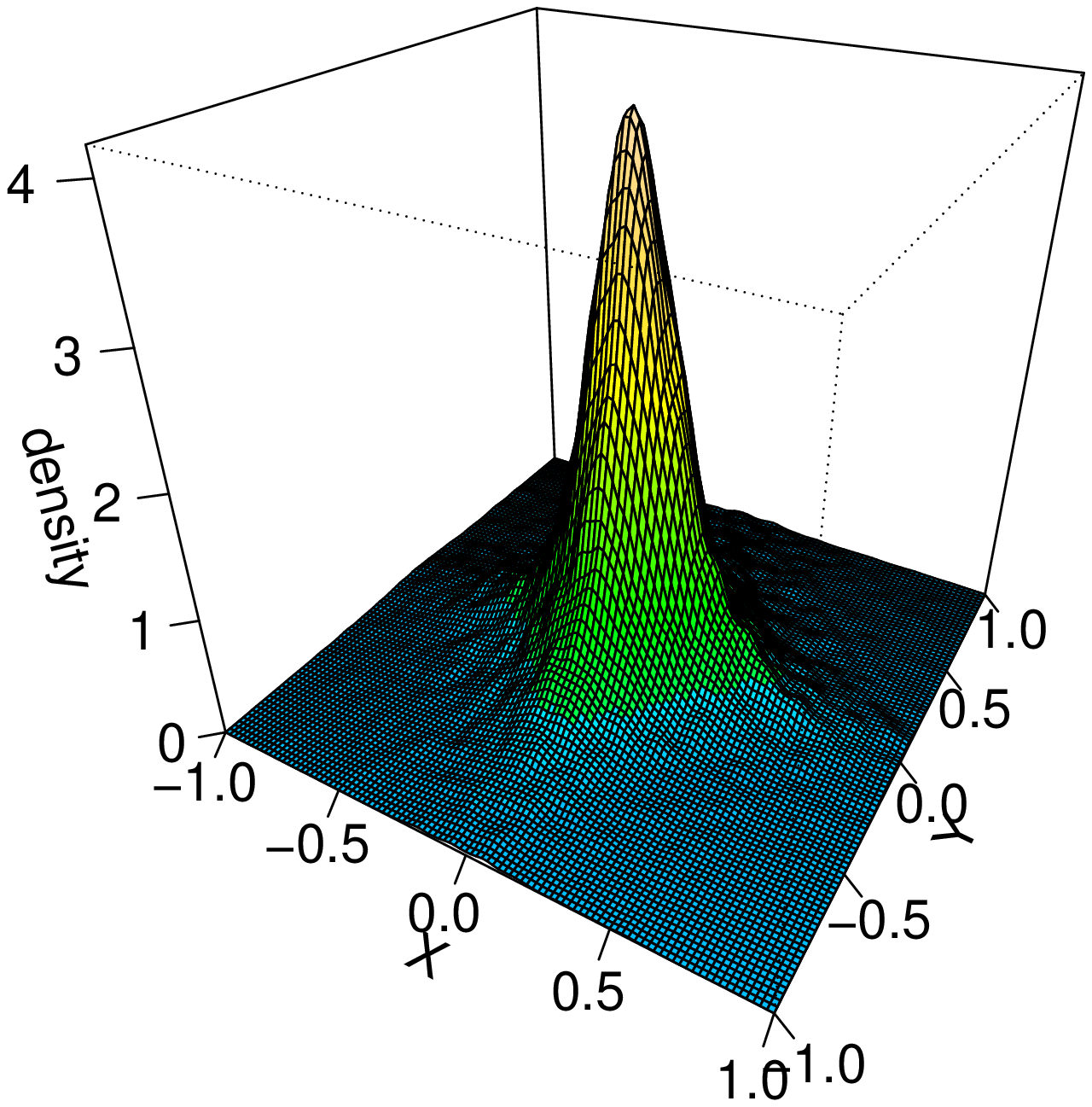}
\includegraphics[width=45mm,height=45mm]{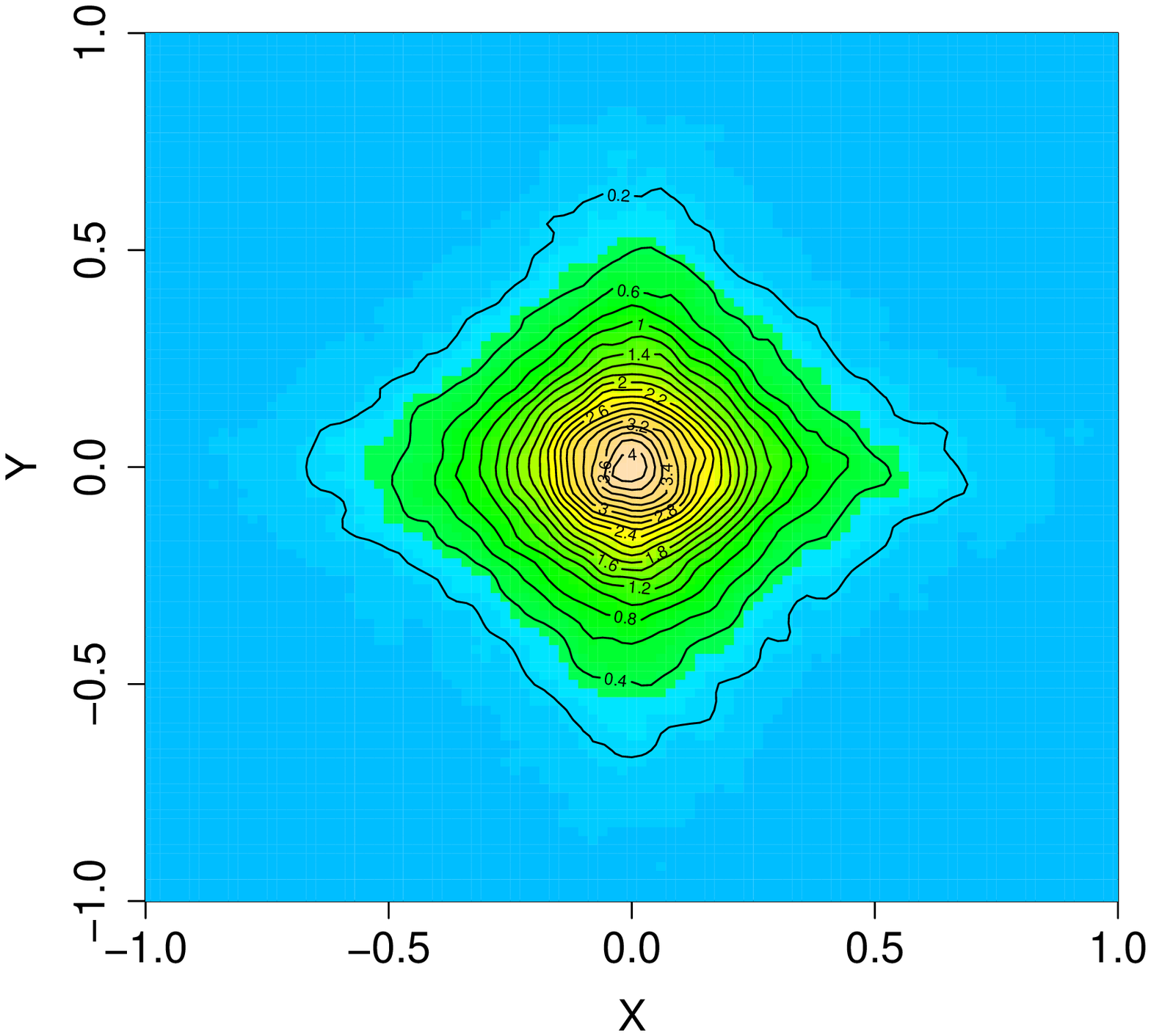}\\
\includegraphics[width=45mm,height=45mm]{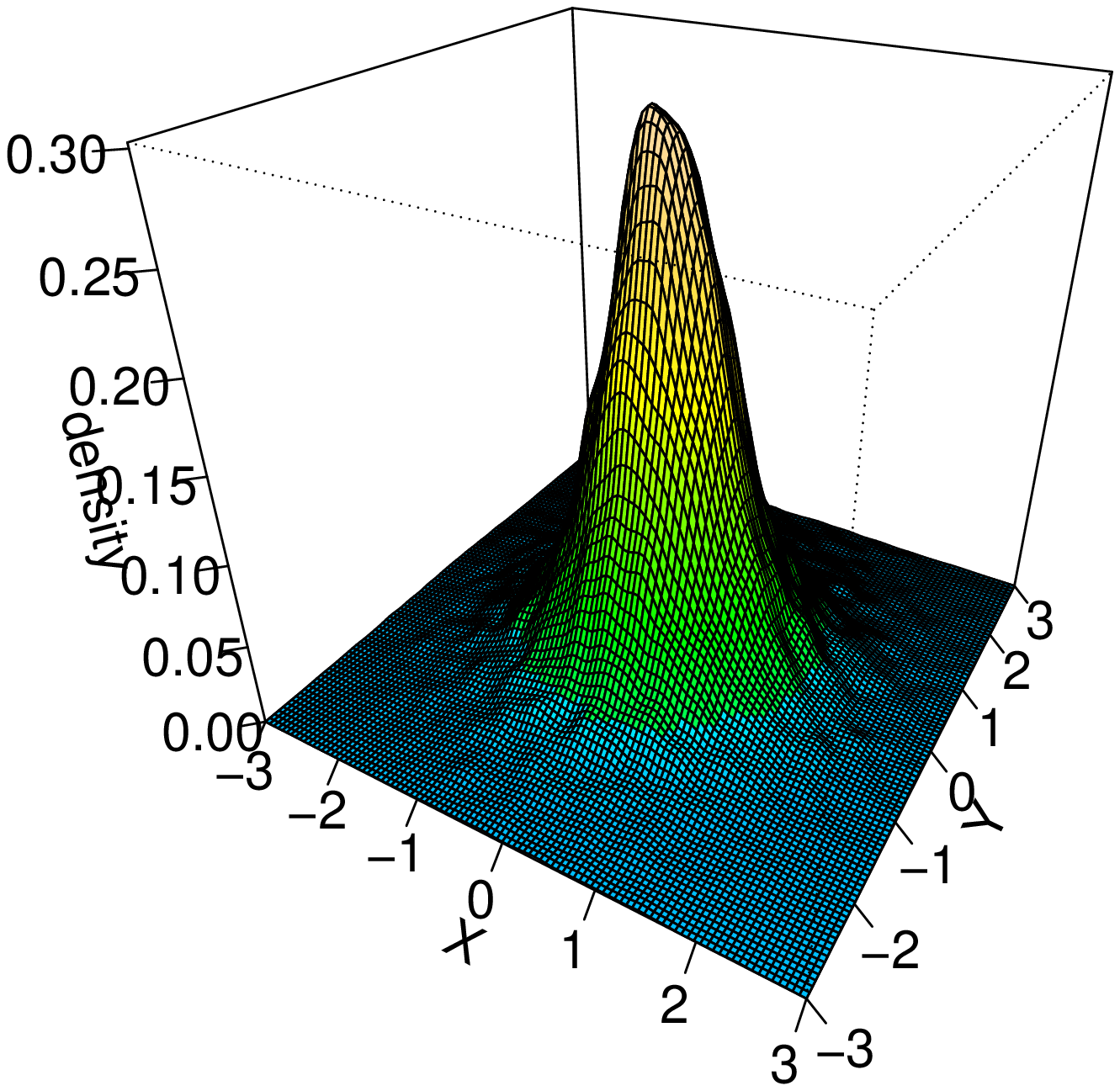}
\includegraphics[width=45mm,height=45mm]{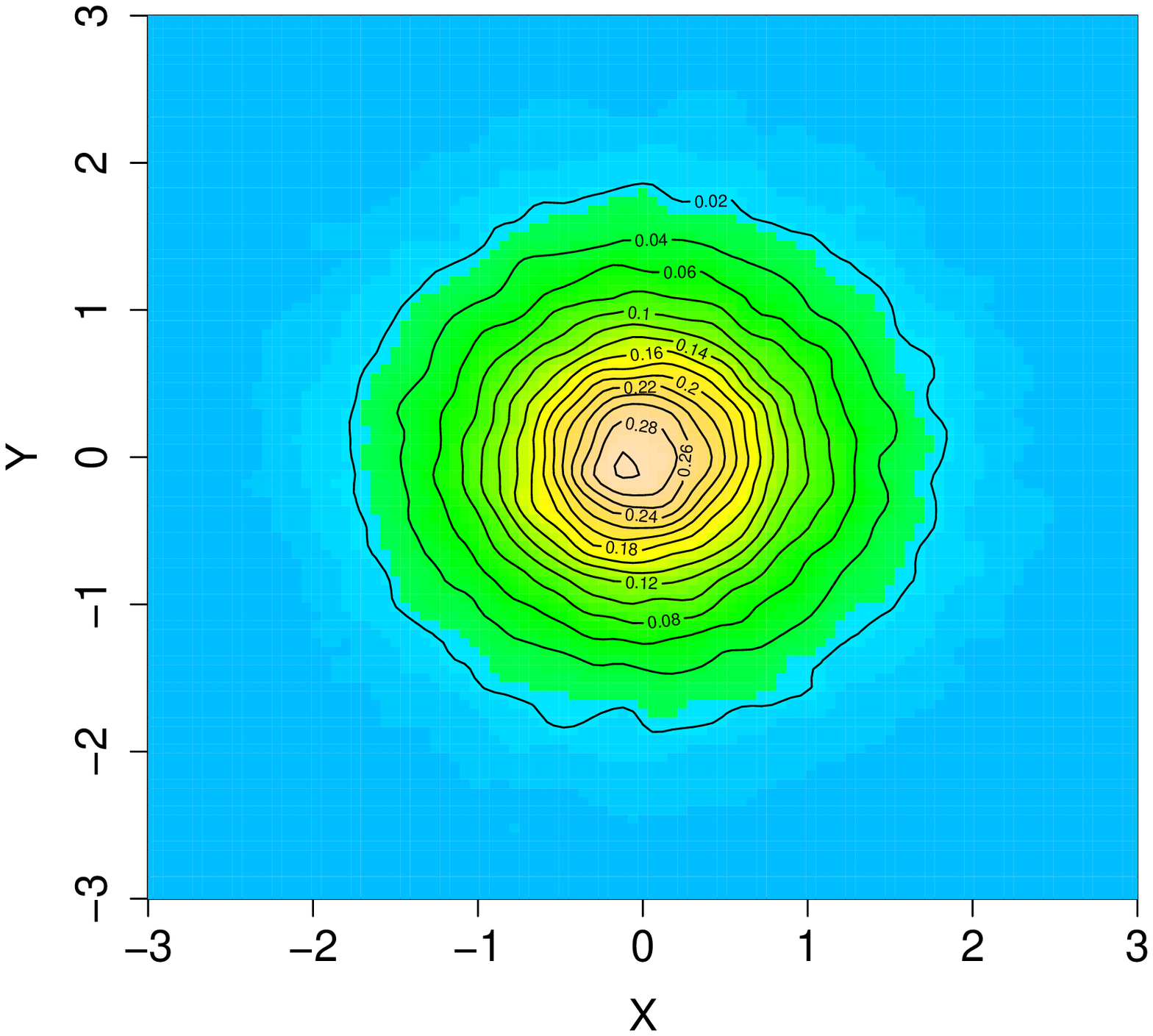}\\
\includegraphics[width=45mm,height=45mm]{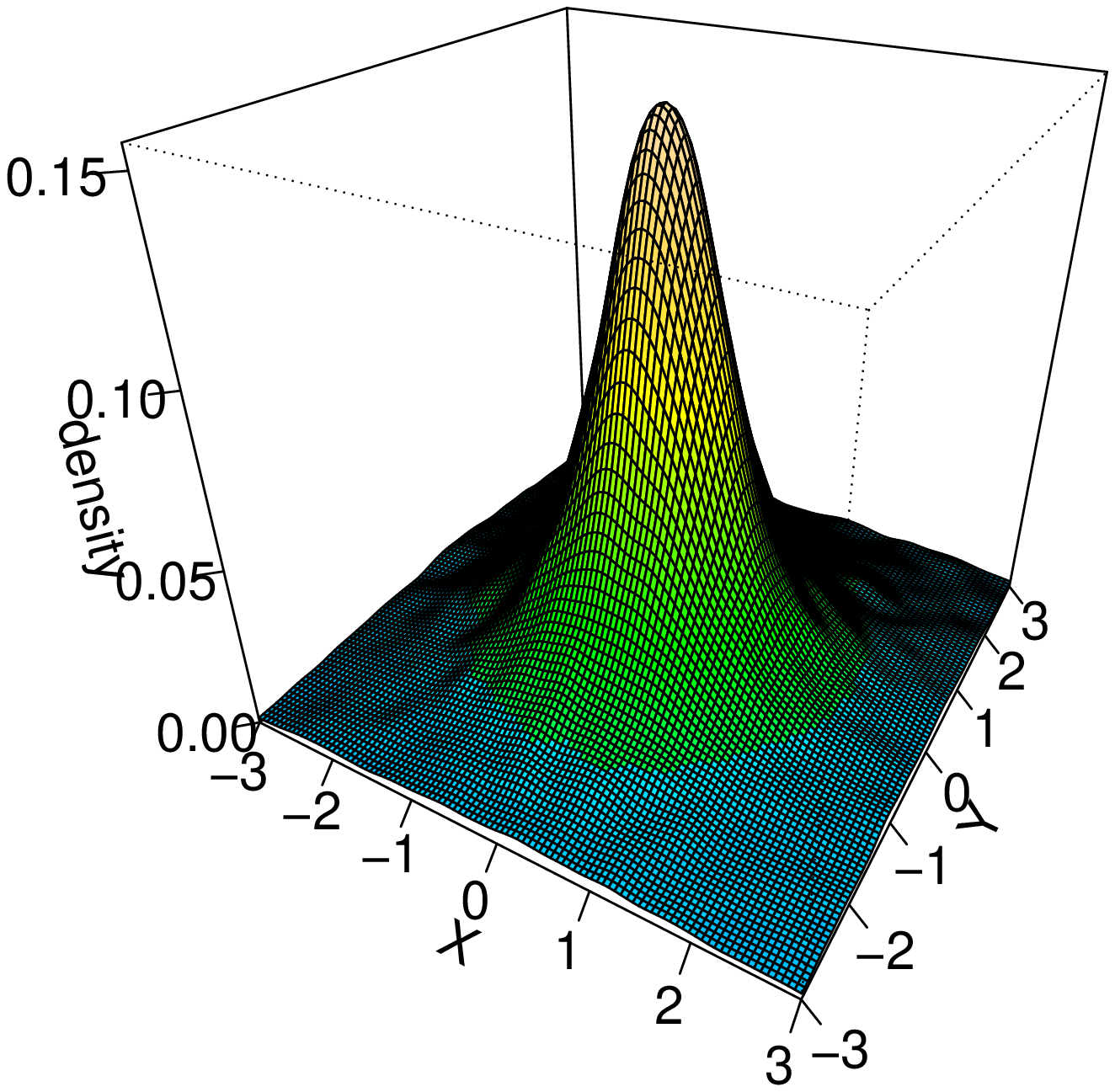}
\includegraphics[width=45mm,height=45mm]{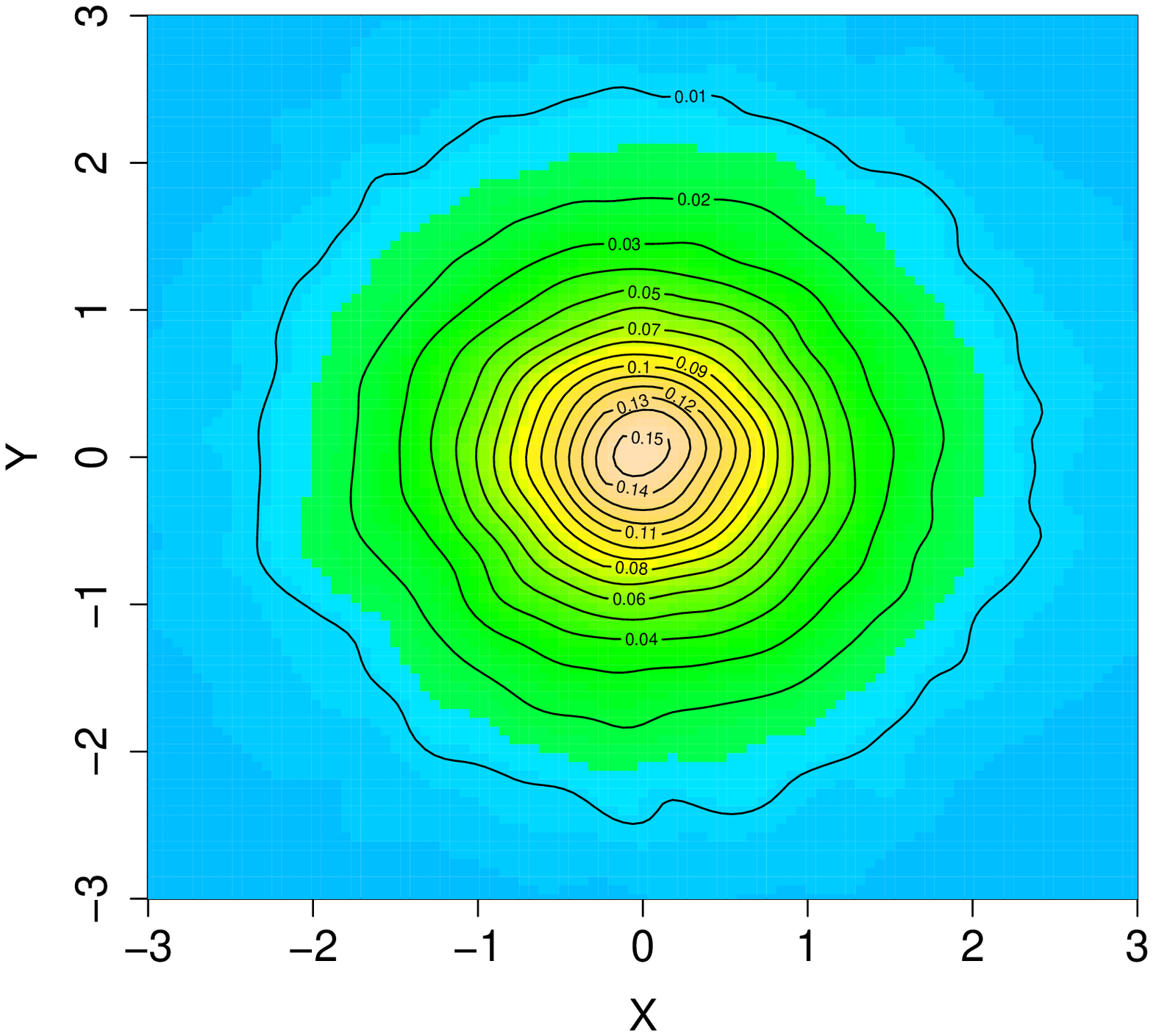}\\
\includegraphics[width=45mm,height=45mm]{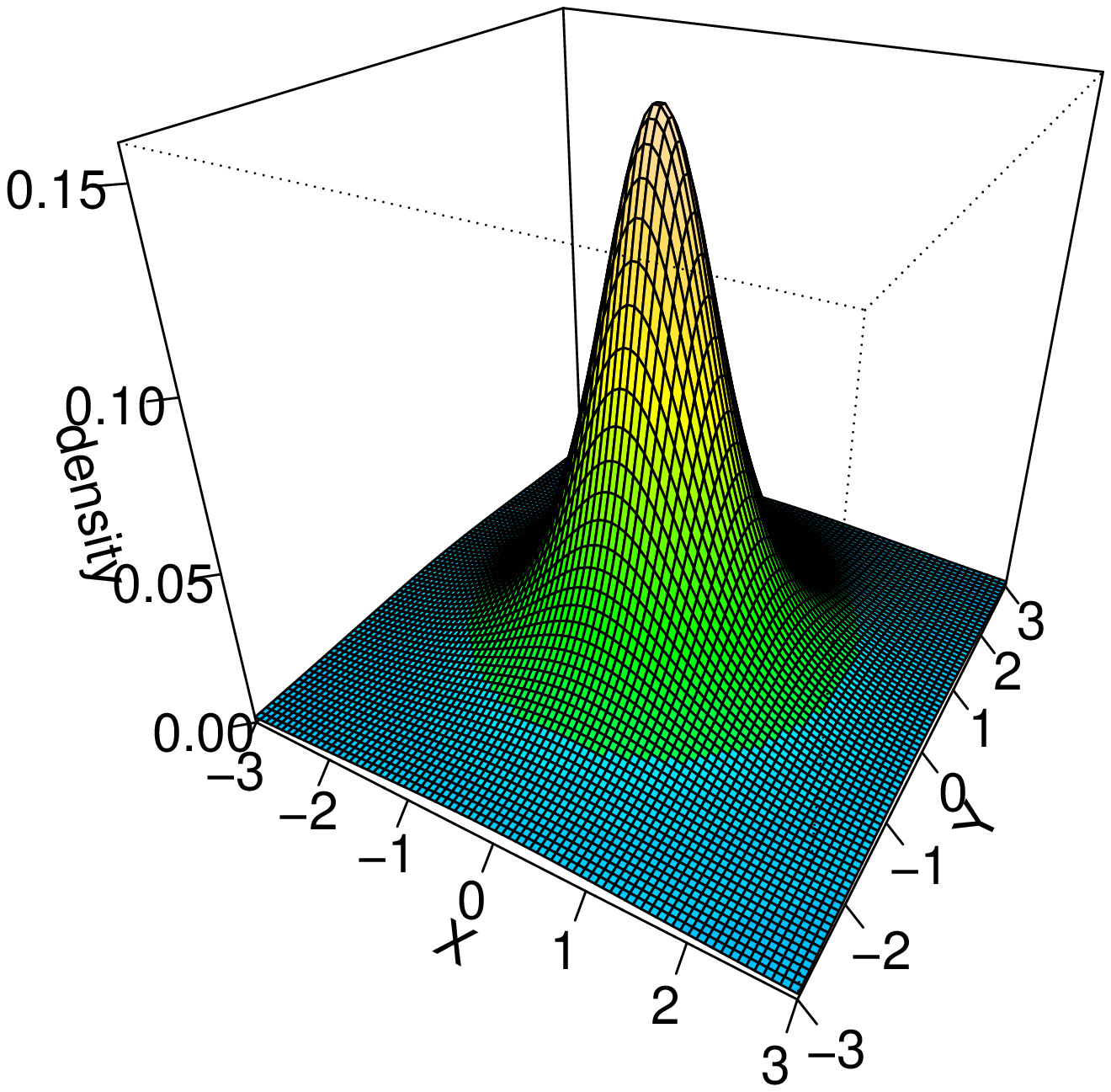}
\includegraphics[width=45mm,height=45mm]{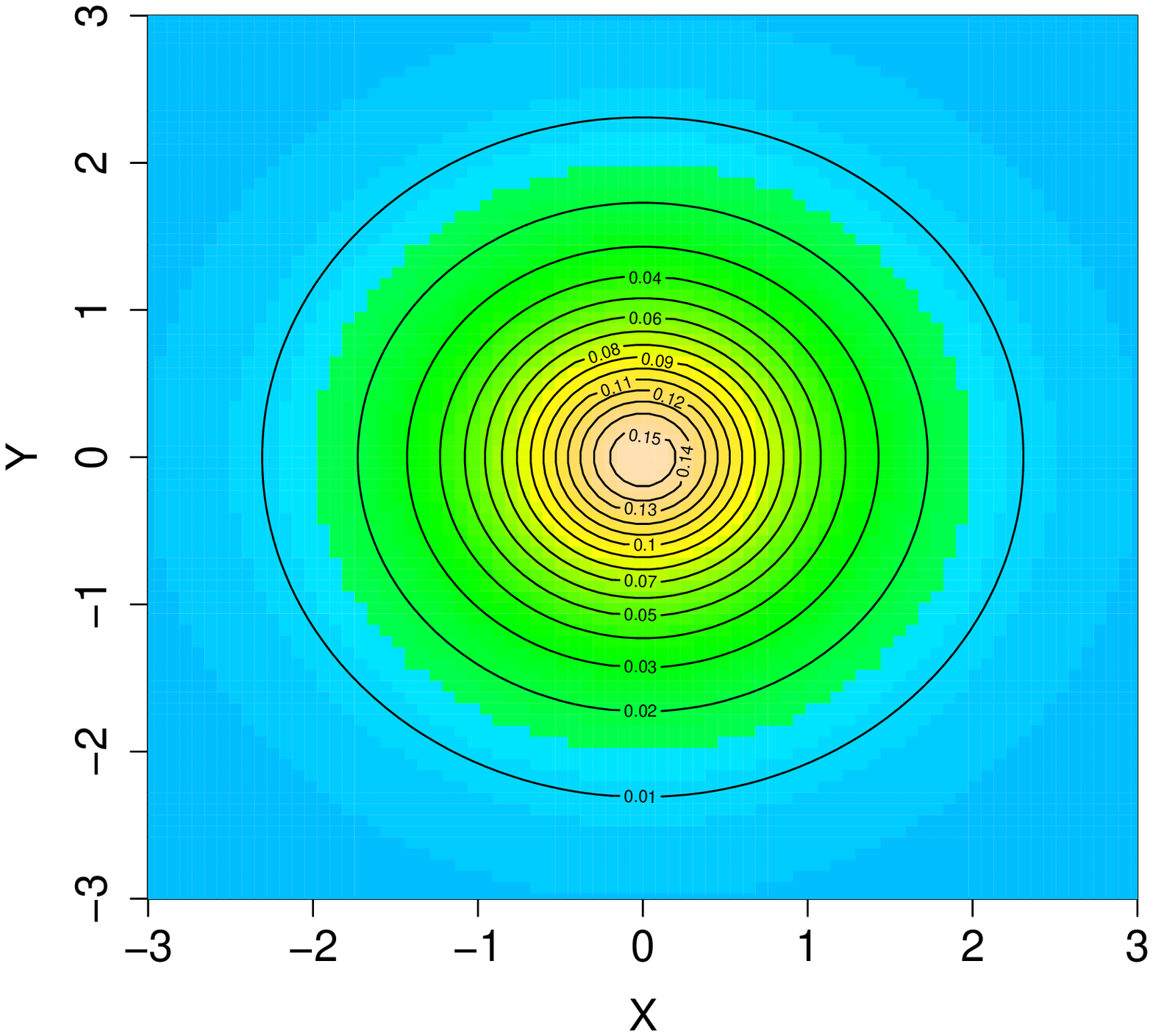}
\caption{ Cauchy target;  surrogate pdf evolution  inferred from  $100 000$ sample trajectories and the   OXY  projection  for running times  a) $t=0.2$, b) $t=3$, c) $t=500$.   A subfigure  d)
refers to the target Cauchy pdf  \ref{l13}).  In the course of simulations all trajectories  were started from  $(0,0)$.}
\end{center}
\end{figure}

It is clear that  the   surrogate pdf evolution consistently goes towards an invariant asymptotic pdf   (compare subfigures  c)   and  d)).   We note a lowering and   flattening of the maximum around  $(0,0)$,
  in consistency with the ultimate target outcome, whose height is  $1/\pi$ as  follows from  $\rho_*(0,0)=1/\pi$.   Visually  accessible in homogeneities   of  circular shapes in  OXY projections  b) i c),  are
   a consequence of  still relatively low number (100 000)  of sample paths data  and approximations involved in evaluating involved integrals.   The shape of OXY projection   a)  is a consequence of the initial data choice.
   Our problem has  a radial symmetry. Therefore in Fig. 2 we depict  a projection of the trajectory  induced  data  upon the $y=0$ plane.   The projection shows as well a consistent convergence towards  the target pdf.

An additional control method   for the path-wise   inferred pdf evolution,   addresses the  time evolution    and an asymptotic behavior of    the pdf  moments  $<R(t)>$  and  $<R^2(t)>$.  Here  $<R(t)>$ is the mean  distance of  $(x,y)$ points of a  trajectory
form the origin  $(0,0)$ at the running time instant $t$, while  $<R^2(t)>$   is a mean square  distance from  $(0,0)$.  In view of
 \be
\begin{split}
<R>_{as}=\frac{1}{\pi}\iint\limits_{\mathbb{R}^2}\sqrt{x^2+y^2}e^{-x^2-y^2}=\frac{\sqrt{\pi}}{2}\thickapprox 0.886,\\
<R^2>_{as}=\frac{1}{\pi}\iint\limits_{\mathbb{R}^2}(x^2+y^2)e^{-x^2-y^2}=1,\label{l12}
\end{split}
\ee
the   $<R(t)>$ dynamics should set down at  $\sqrt{\pi}/2$,  while this of  $<R^2(t)>$   at $1$.  Fig. 3 depicts the evolution  of    $<R(t)>$ and  $<R^2(t)>$, inferred from the simulated sample of $100 000$  jumping paths.

The observed convergence $<R(t)>  \rightarrow 1/\pi$  and $<R^2(t)> \rightarrow 1$ validates the number generator choice, we have used  to arrive  at   sample   jumping paths.

\begin{figure}[h]
\begin{center}
\centering
\includegraphics[width=45mm,height=45mm]{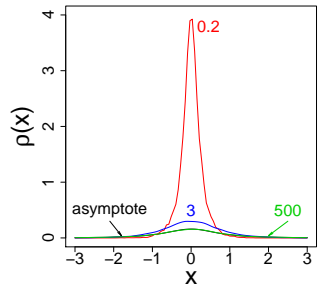}\\
\caption{ The projection of the previous pdf data      on the $y=0$ plane,   at   time   instants  a) $t=0.2$, b) $t=3$, c) $t=500$, d) target pdf projection.}
\end{center}
\end{figure}

\subsection{Logarithmic confinement: 2D   Cauchy  target.}

We consider the  target pdf $\rho _*$ in the  2D  Cauchy form:
\be
\rho_*(x,y)=\frac{1}{2\pi}\frac{1}{(1+x^2+y^2)^{3/2}}.\label{l13}
\ee
Like previously, we take the Cauchy driver, $\mu=1$,   as a reference L\'{e}vy stable noise. Accordingly:
\be
C_1w_\phi(\xi+x,\eta+y|x,y)=\frac{1}{2\pi}\frac{1}{(\xi^2+\eta^2)^{3/2}}\left(\frac{1+x^2+y^2}{1+(x+\xi)^2+(y+\eta)^2}\right)^{3/4}.\label{l14}
\ee
Proceeding like in the Gaussian case,  for small  $\xi$  and  $\eta$,    Eq.  (\ref{l14}) can be approximated by
\be
\frac{1}{2\pi}\frac{1}{(\xi^2+\eta^2)^{3/2}}\left(1-\frac{3x}{2(1+x^2+y^2)}\,\xi-\frac{3y}{2(1+x^2+y^2)}\,\eta\right),\label{l15}
\ee
where terms linear in $\xi$ i $\eta$ were preserved.  The result can be analytically integrated term after term  by employing Eqs. (17)..

Accumulated  trajectory data have been analyzed to produce Fig.  4, where a surrogate pdf   evolution is displayed.  Fig. 5  reproduces the $y=0$ projection   of the obtained pdf data.   If compared with the 1D case analyzed in Ref. \cite{zaba}
 a convergence to Boltzmannian equilibrium (target pdf)   is   substantially slowed down.

\begin{figure}[h]
\begin{center}
\centering
\includegraphics[width=45mm,height=45mm]{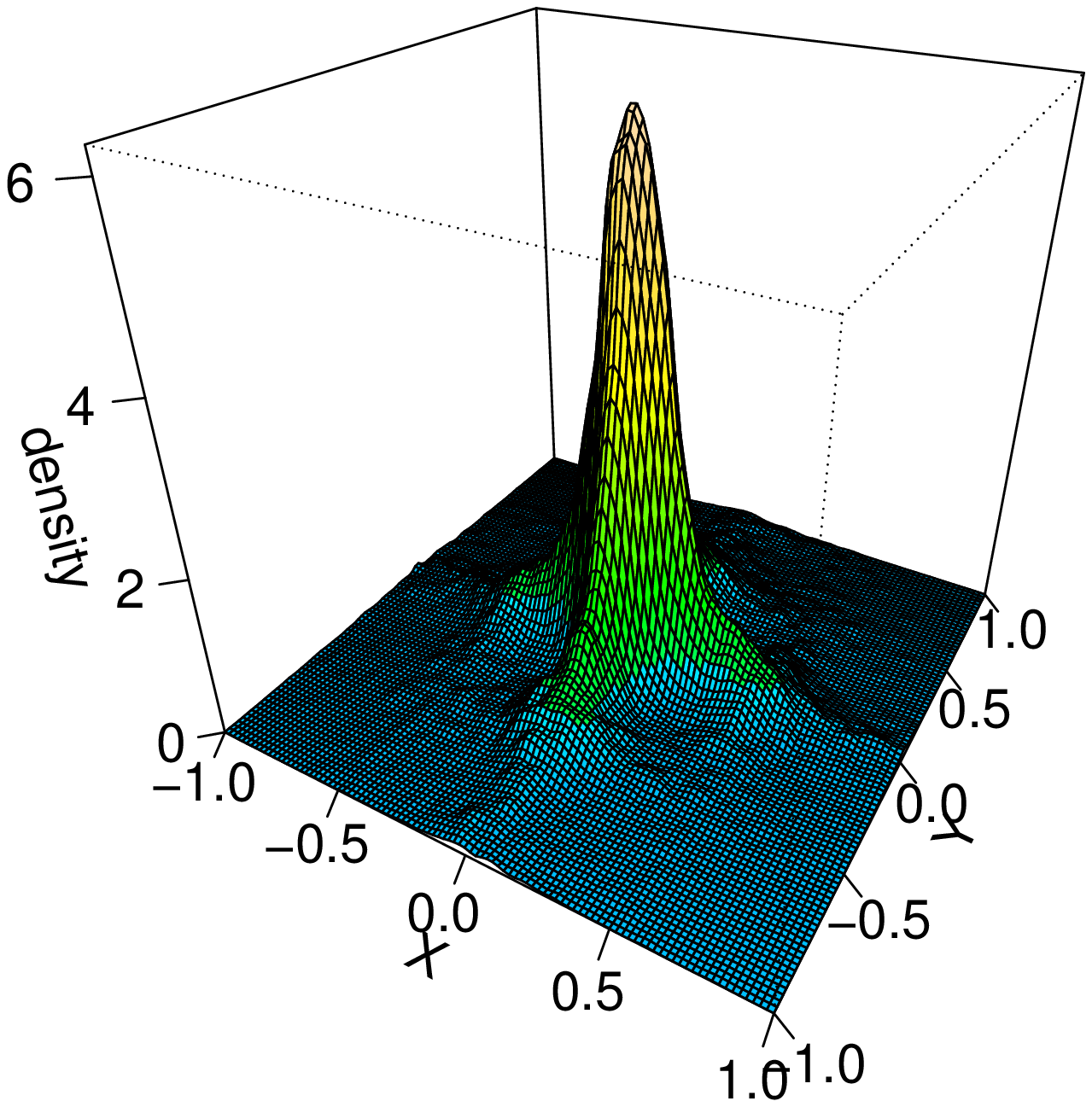}
\includegraphics[width=45mm,height=45mm]{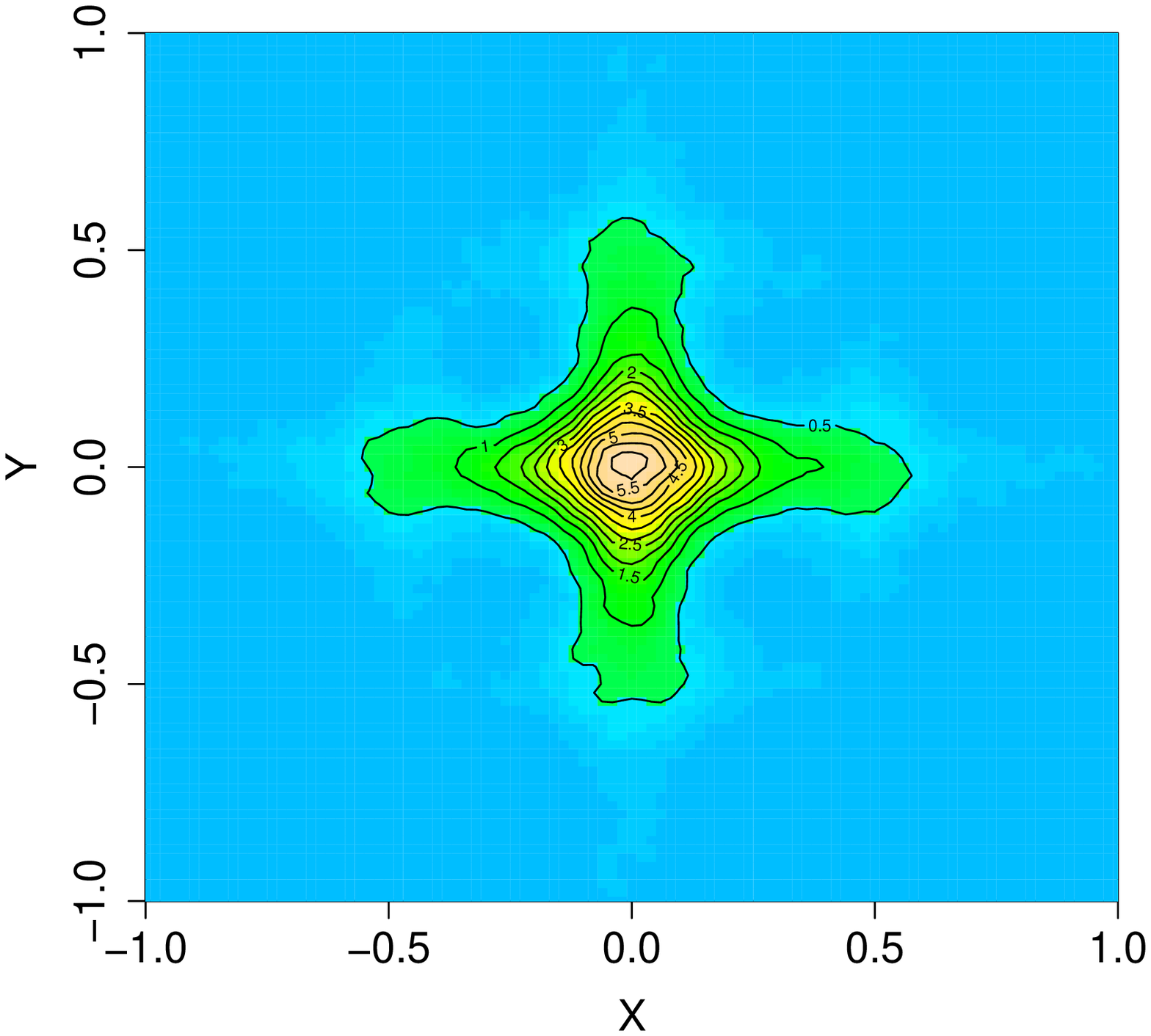}\\
\includegraphics[width=45mm,height=45mm]{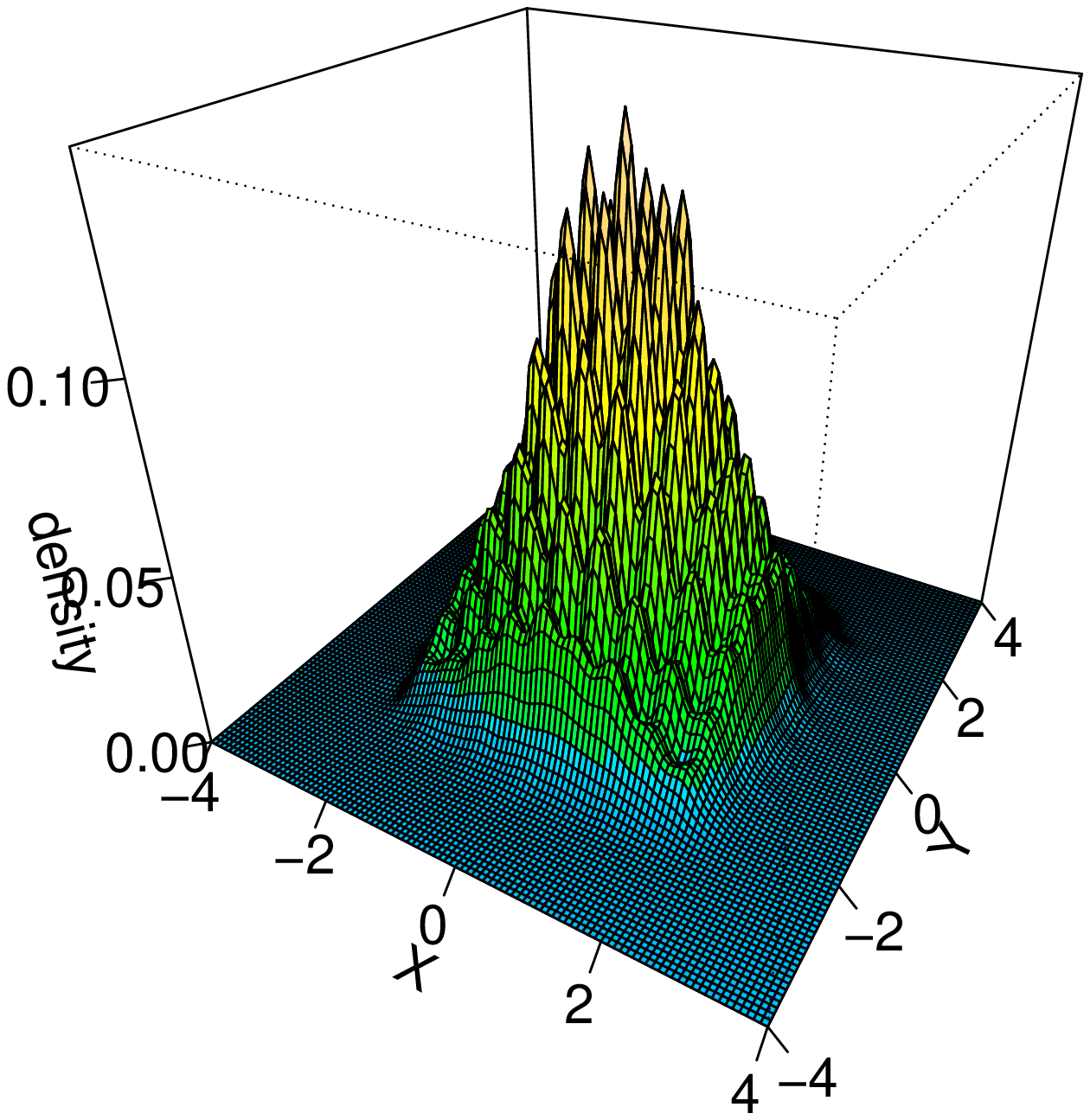}
\includegraphics[width=45mm,height=45mm]{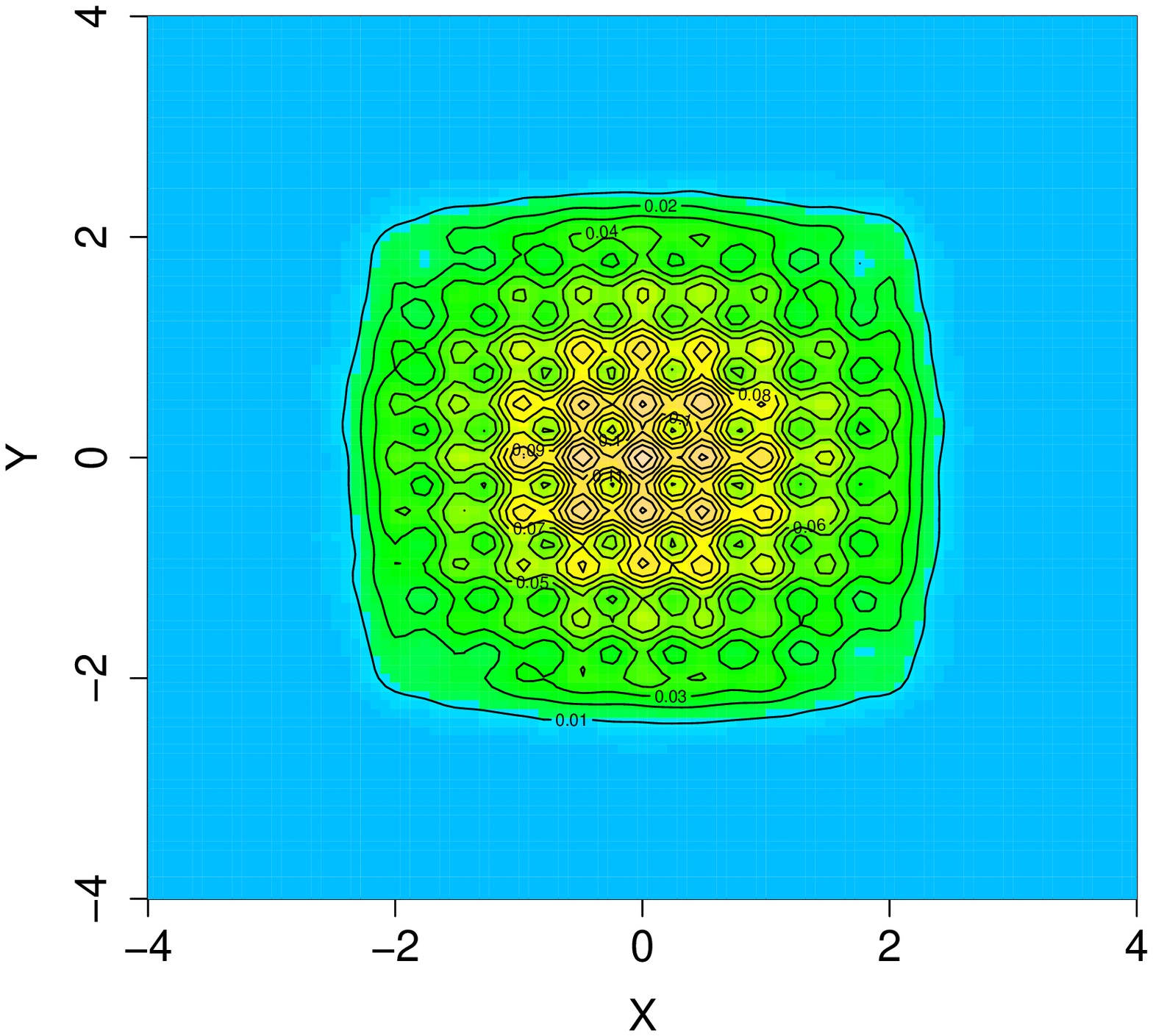}\\
\includegraphics[width=45mm,height=45mm]{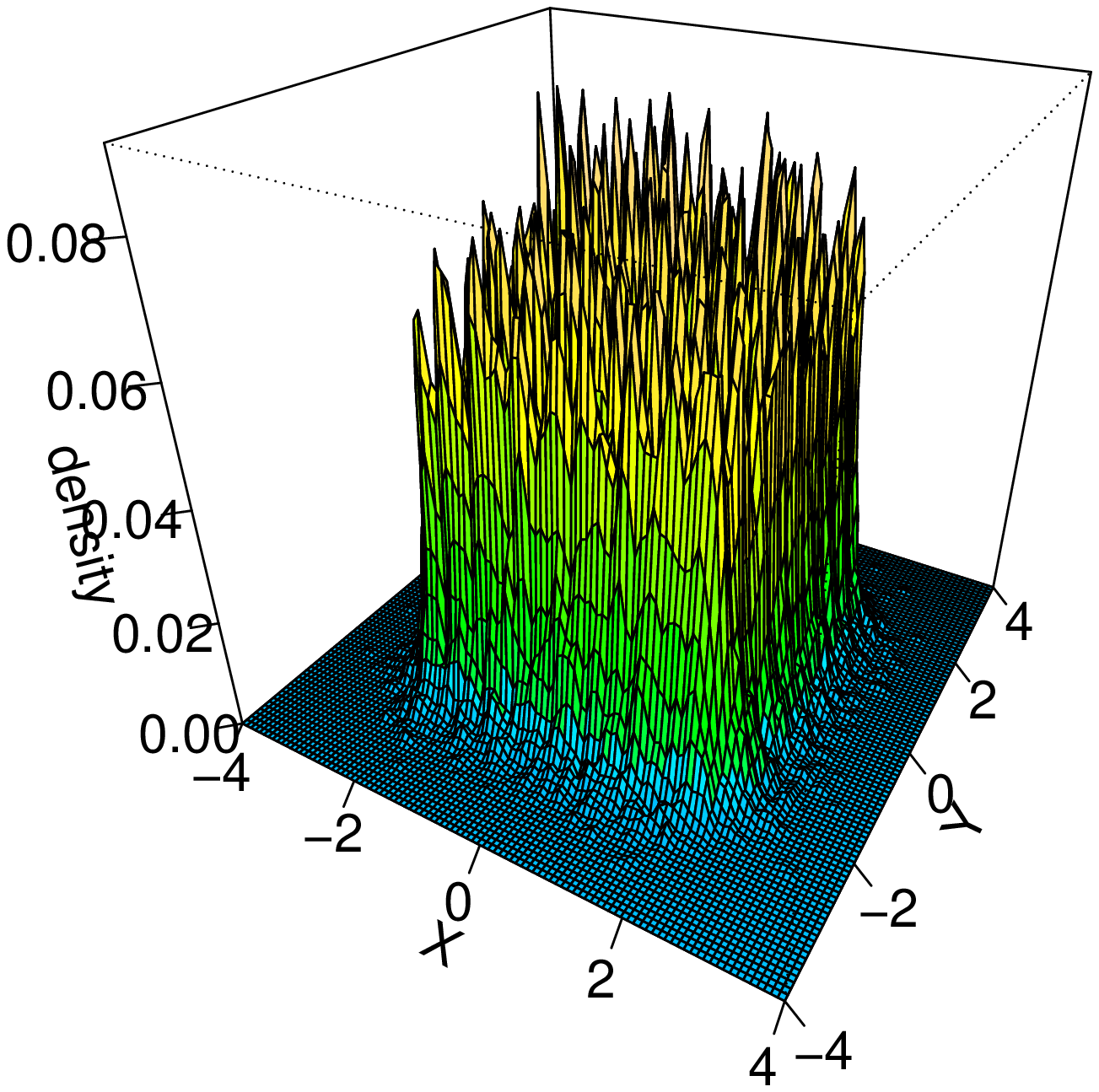}
\includegraphics[width=45mm,height=45mm]{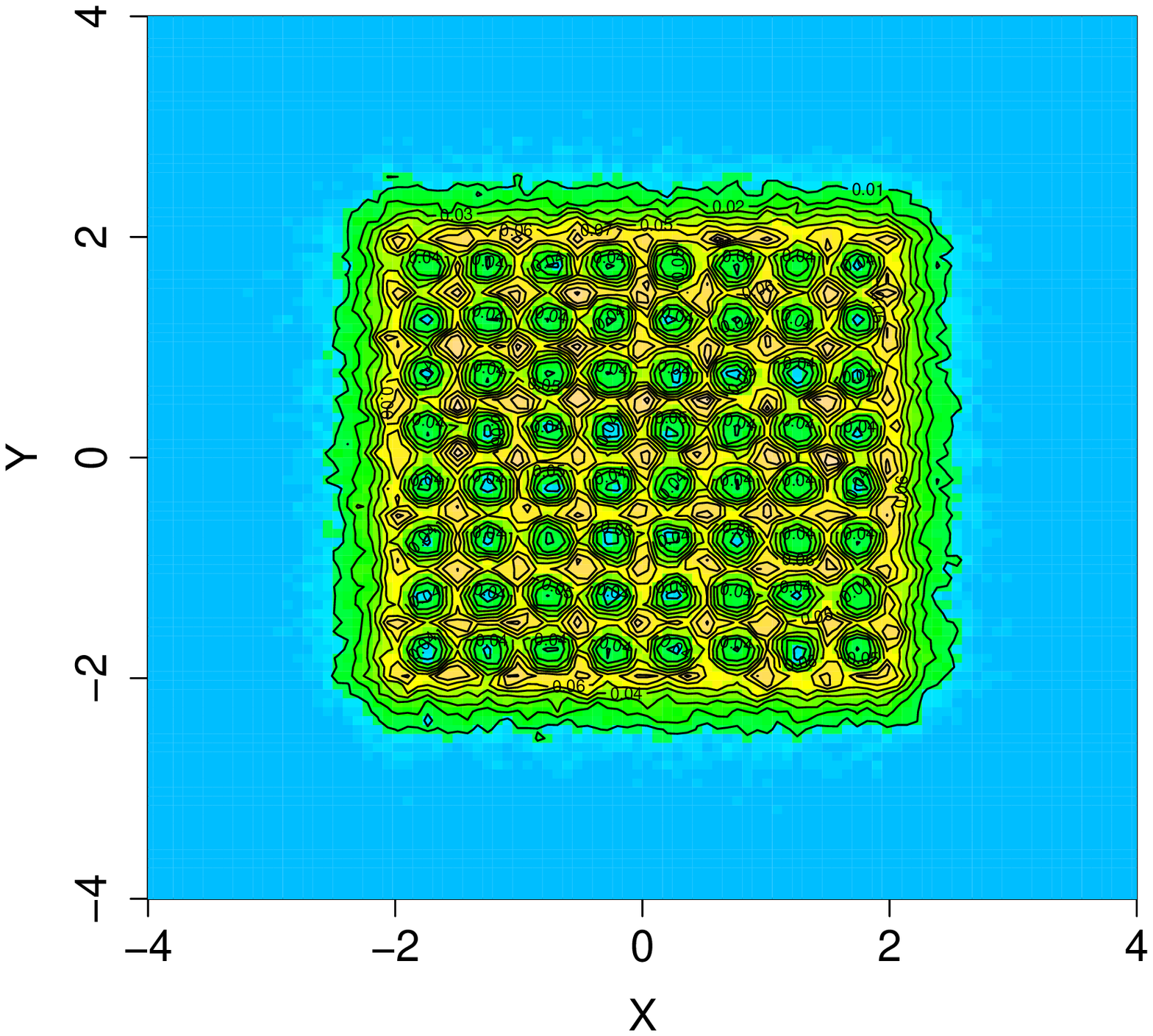}\\
\includegraphics[width=45mm,height=45mm]{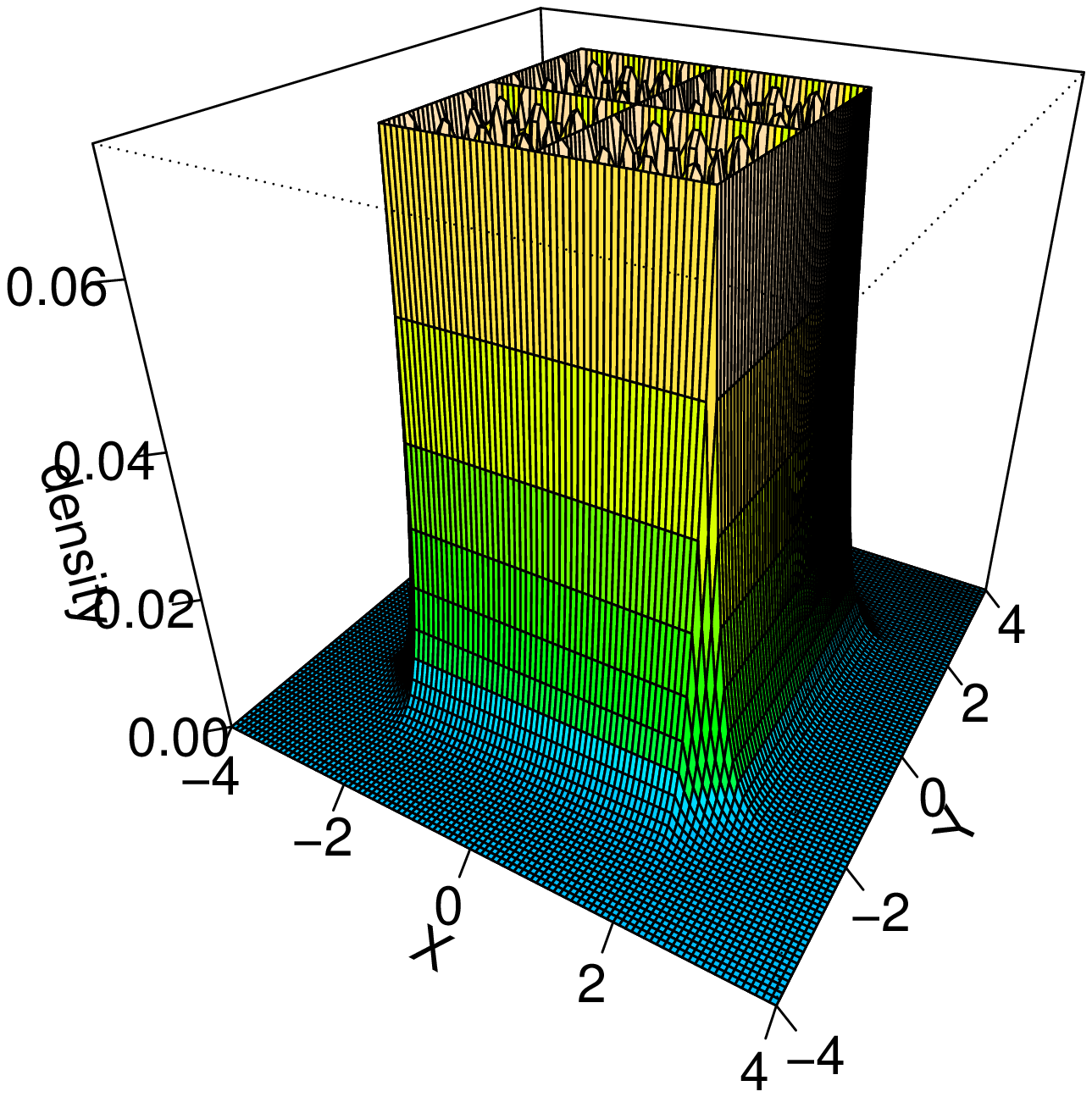}
\includegraphics[width=45mm,height=45mm]{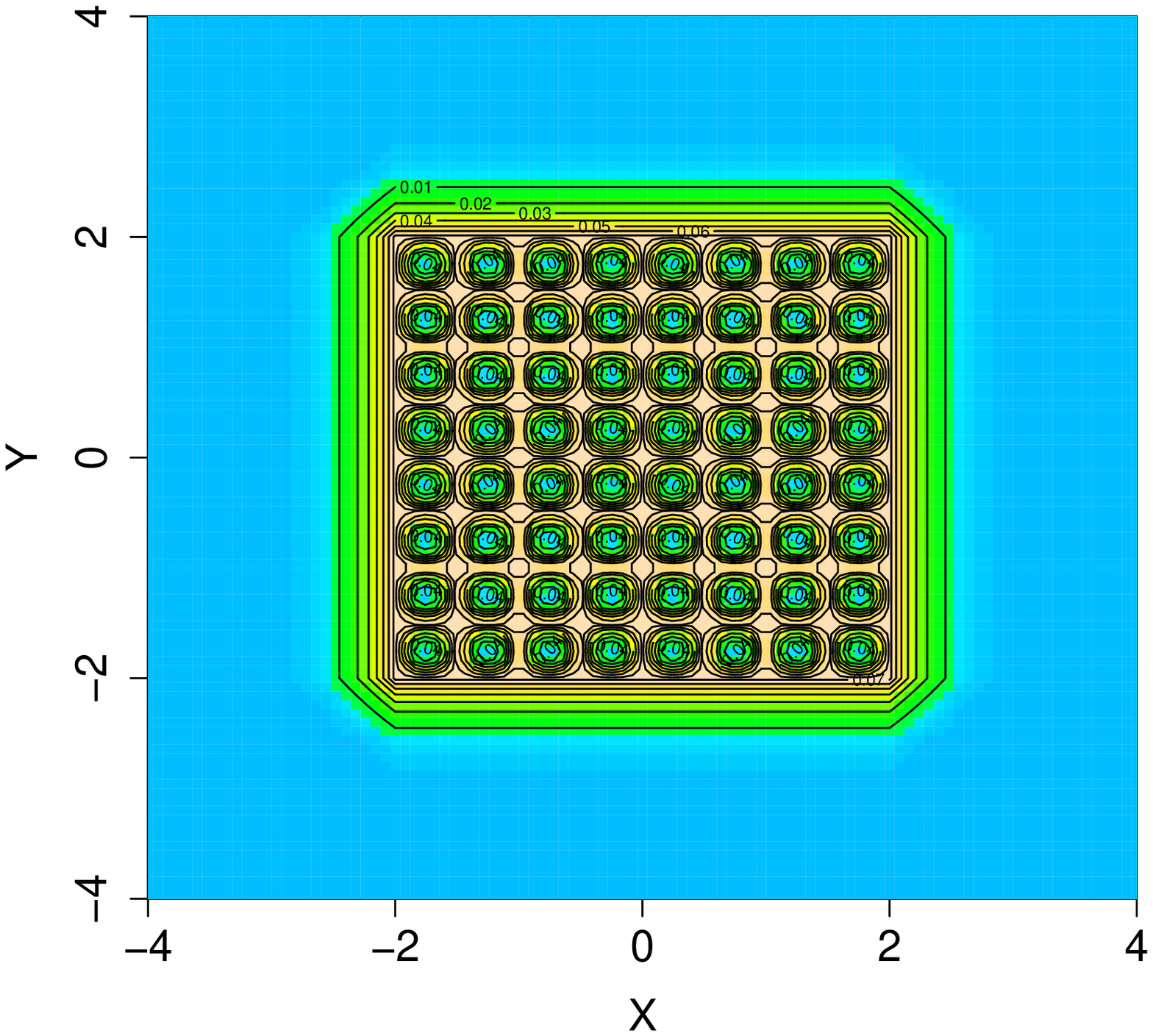}
\caption{Locally periodic target;  surrogate pdf evolution and its OXY projection inferred from $100 000$  trajectories  at running time instants  a) $t=0.2$, b) $t=3$, c) $t=300$.   The subfigure d)  depicts the asymptotic pdf  (\ref{l16}).
All trajectories have been started from  $(0,0)$.}
\end{center}
\end{figure}

\subsection{Locally periodic confinement  in  $ R^2$.}
\begin{figure}[h]
\begin{center}
\centering
\includegraphics[width=45mm,height=45mm]{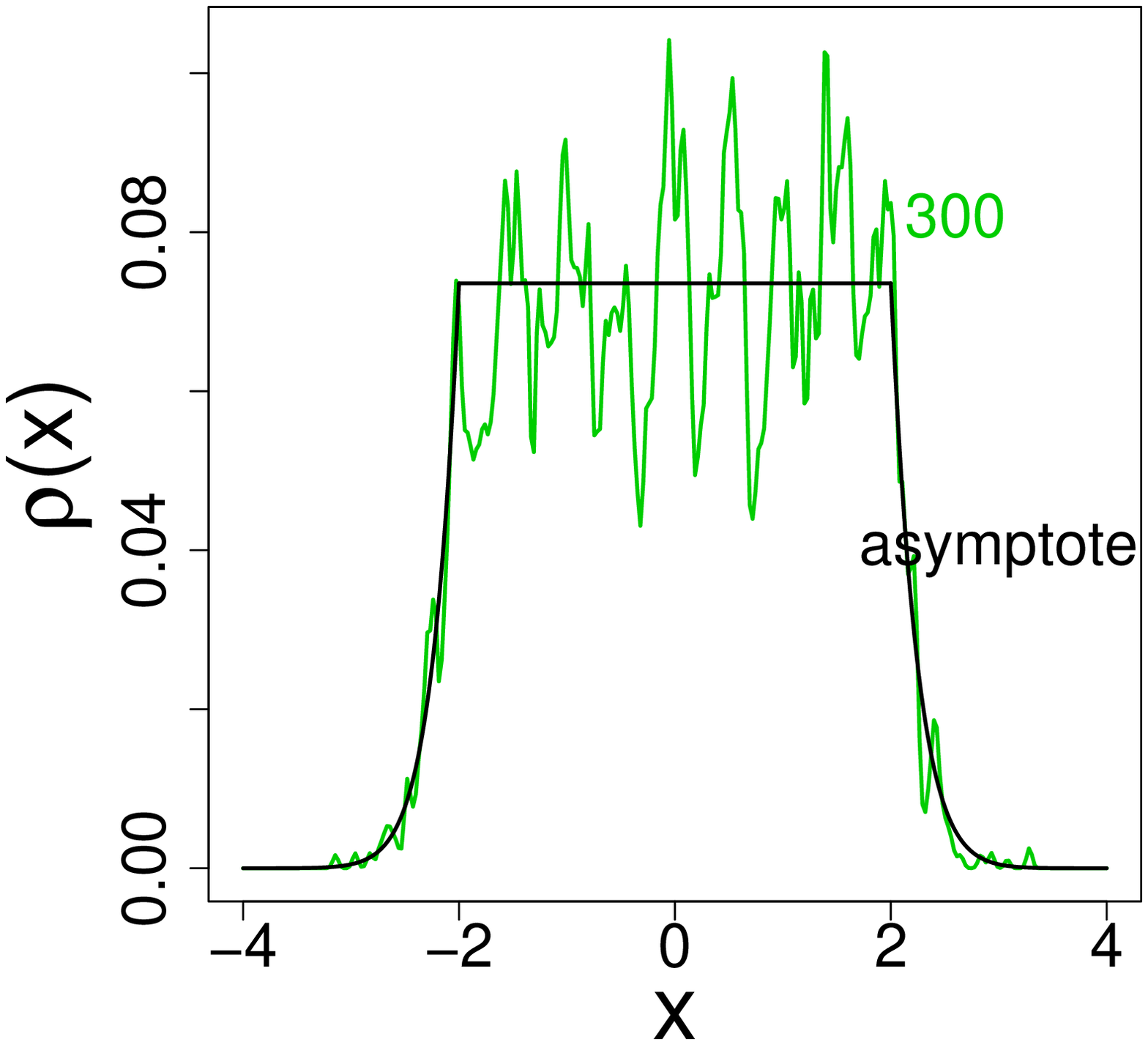}\\
\includegraphics[width=45mm,height=45mm]{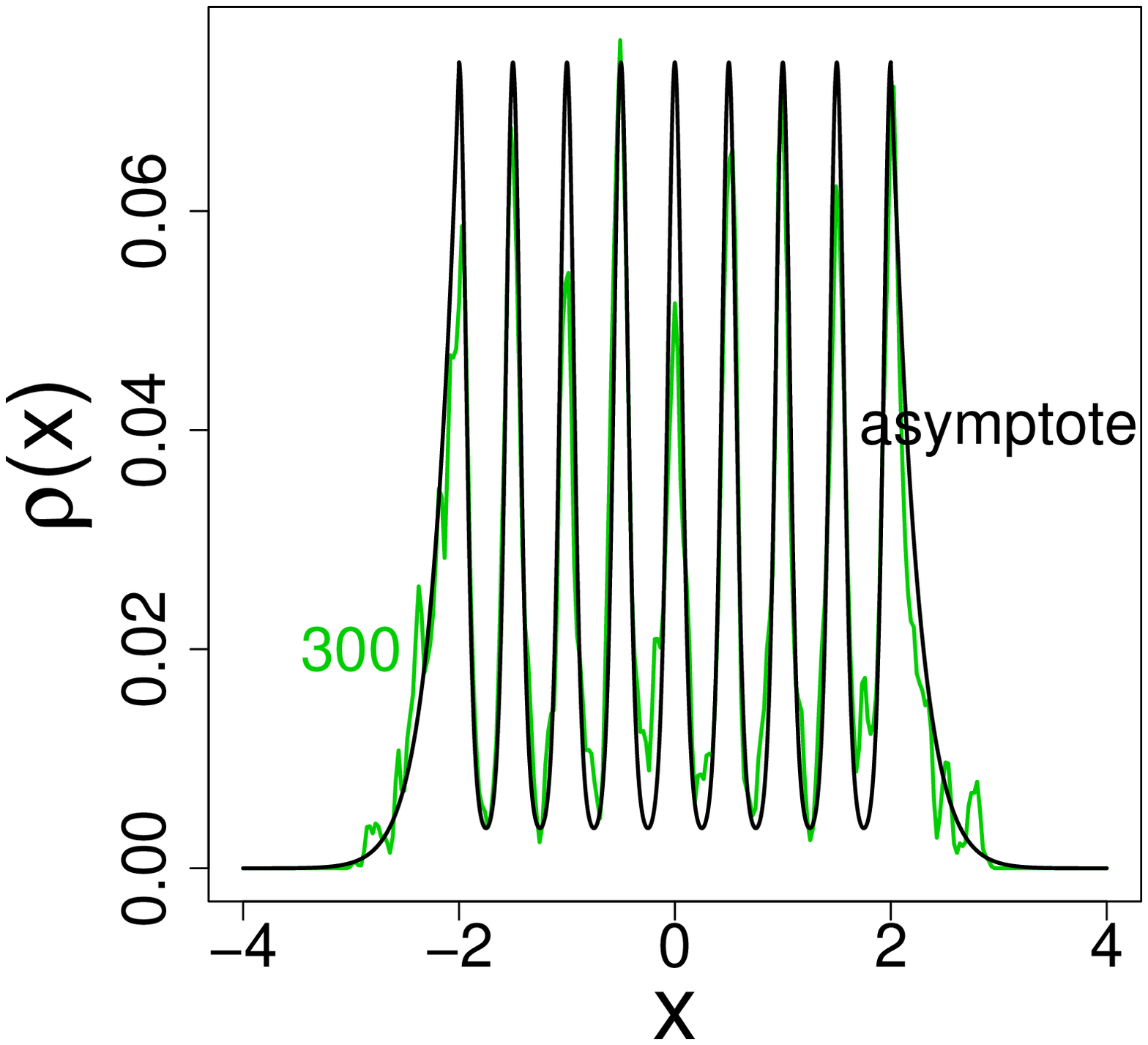}\\
\includegraphics[width=45mm,height=45mm]{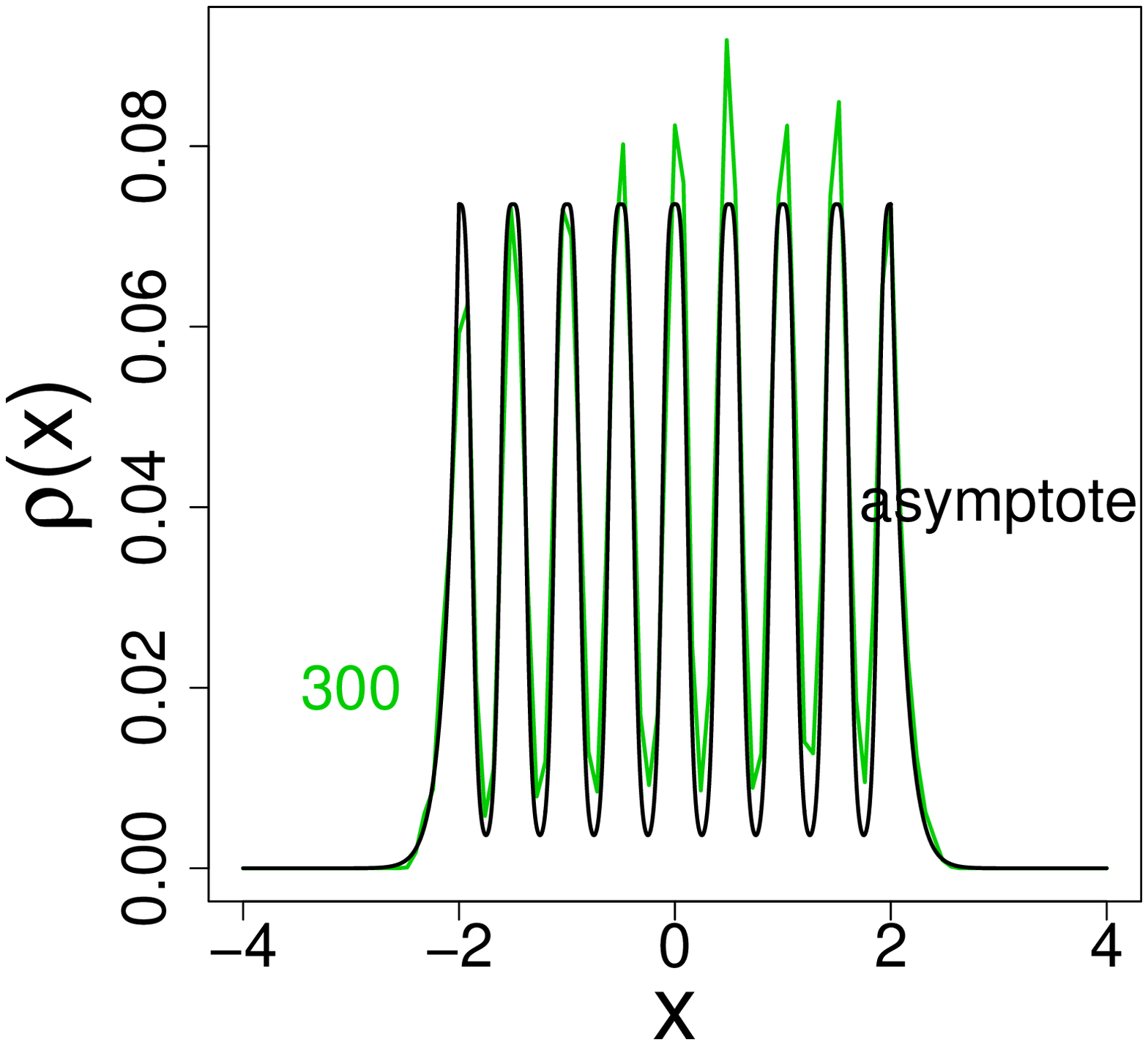}\\
\caption{ Projections of the surrogate pdf data at $t=300$  upon planes  a) $y=0$, b) $y=1/4$, c) $y=x$.  An "asymptote" refers  to the target pdf projections}
\end{center}
\end{figure}

We consider   target pdf whose   Boltzmanian   exponent    is locally periodic (within a finite rectangle) and    \it almost \rm   entirely     localized   within   a      finite spatial area
due to  harmonically confining tails  of the potential:
\be
\rho_*(x,y)=\frac{1}{C}\left\{
                         \begin{array}{ll}
                           e^{-3\sin^2(2\pi x)\sin^2(2\pi y)}, & \hbox{$|x|\leqslant 2$ i $|y|\leqslant 2$;} \\
                           e^{-x^2-y^2+8}, & \hbox{$|x|> 2$ i $|y|> 2$;} \\
                           e^{-x^2+4}, & \hbox{$|x|> 2$ i $|y|\leqslant 2$;} \\
                           e^{-y^2+4}, & \hbox{$|x|\leqslant 2$ i $|y|> 2$,}
                         \end{array}
                       \right.
\label{l16}
\ee
The normalization  constant  $C$   actually reads  $C=13.5921$.

Subsequently   adopted  numerical integration  routines  heavily   rely   on   the  experience gained during   our    previous case studies.   In Fig. 6 we report the surrogate pdf evolution, inferred from $100 000$ sample trajectories.
 A convergence  rate  to the asymptotic (target)  pdf   is satisfactory,  although  a reasonable agreement with the target  data   has been  achieved     for  relatively large running time values,   here  $t=300$.

We  are aware  of the fact  that the number of $100 000$ trajectories may be considered as too small and not sufficiently representative sample.  Our tentative $300 000$ paths  data do not show  significant qualitative changes in the obtained evolution picture.

We should mention that  there are significant statistical fluctuations  to be kept under control.   They become  are very conspicuous if the number of involved trajectories gets  significantly lowered by imposing  constraints
(like e.g. various spatial  projections).
All l trajectory data,  after being gathered, are   safely  stored in  the computer memory. Therefore  we can   get access to  any conceivable and   more detailed statistical  picture  of what is going on, even if  the outcome  is  hampered by
 significant   random deviations from the reference (target) pdf data..

A sample of   such  fluctuating  data is provided in Fig. 7, where we have considered   projections of the  surrogate pdf data  upon planes $y=0$, $y=1/4$   and $y=x$   at time $t=300$. We have set  them in a direct comparison with respective
target pdf  (\ref{l6})  data.

\section{Outlook}

We have  taken  into consideration    jump-type processes which can  not be handled  by standard  stochastic differential equation methods
 (e.g. the  Langevin  modeling,   where  a conspicuous   motion "tendency"  quantified by an additive   drift  term,    can be  unambiguously   isolated from  the noise contribution).
  Existing popular algorithms cannot   provide a  direct numerical simulation  of  sample paths of such non-standard processes.
    In the present paper, we    have   proposed    a working method to generate stochastic trajectories
(sample paths) of a random jump-type process   that avoids any reference to a stochastic
differential equation. An additional gain of that procedure is that we are in fact capable of reliably approximating the time evolution $\rho (x,t)$ of a true (typically not available in a closed analytic form) solution of the master equation.

 To this end we have modified the Gillespie algorithm,   \cite{gillespie,gillespie1}, normally devised for sample paths
generation if the transition rates refer to a finite number of states of a system.
The essence of our modification is that we take into account the continuum of possible transition rates, thereby
changing the finite sums in the original Gillespie algorithm into integrals. The corresponding procedures for stochastic
trajectories generation have   been changed accordingly.

 In other words, here we  are able   (i)    to extract  the background sample paths
of a jump process  and   (ii) to   infer a   reliable approximation of an actual   (analytically unavailable)  solution of  the  master  equation  (7)- (8).
We  emphasize once more here, that we have focused on those  jump-type processes that cannot be modeled
by any stochastic differential equation of the Langevin type.

Although heavy-tailed  L\'{e}vy stable drivers were involved in the present considerations, we have clearly confirmed
that a large  variety of stationary target distributions is dynamically accessible   for  each particular  $\mu \in  (0, 2)$
L\'{e}vy driver choice.  That   variety  comprises not only  the  standard Gaussian pdf, casually discussed in relation
 to the Brownian motion (e.g.  the Wiener process), but the whole non-Gaussian family, associated with the
  L\'{e}vy  stable  conceptual imagery .

 Among heavy-tailed distributions, we have paid attention to the Cauchy pdf which can stand
for an asymptotic target for any  $\mu  \in (0,2) $  driver, provided a steering environment  (e.g. "potential landscape")  is properly devised.
In turn, the   Cauchy driver, while   excited   in  a proper environment,  may lead to an asymptotic pdf with an  arbitrarily large number  of moments, the  previously mentioned  Gaussian case being included.
.
An example of the locally periodic environment has been considered as a toy model for more realistic physical
systems. Our major hunch are strongly inhomogeneous  "potential landscapes",     modeled by  relatively   smooth   potentials. We note that a  radically  extreme variant could be random   potentials   of Ref. \cite{geisel}.

In connection with the  master equation which was our departure point   let us stress that,  even if various mean field data are available in experimentally realizable systems,  it is of    vital   interest to gain    some knowledge about the microscopic dynamics (random paths) realized by the   random system  under consideration.      The detailed analysis of sample path data  with a focus on their specific features like e.g.  ergodicity, mixing or lack of those properties, deserves  a separate analysis.  These goals can be   achieved    as well  within
the present   simulation framework.   It suffices to re-analyze   the   path-wise  data we have  collected  and  stored
  in the trajectory generation process.

\end{document}